\documentclass[twocolumn]{aastex7}
\usepackage{savesym}
\savesymbol{tablenum}
\usepackage{siunitx}
\usepackage{pgf}

\restoresymbol{SIX}{tablenum}

\usepackage{xspace}
\usepackage{physics}

\newcommand\Gaia{\textit{Gaia}\xspace}
\newcommand\rin{r_\mathrm{in}}
\newcommand\rout{r_\mathrm{out}}
\let\vec\mathbf

\begin{document}

\title{Deep Potential: Recovering the gravitational potential and local pattern speed in the solar neighborhood with GDR3 using normalizing flows}

\author[0000-0002-0417-8645]{Taavet Kalda}
\affiliation{Max-Planck-Institut f\"{u}r Astronomie \\
K\"{o}nigstuhl 17 \\
D-69117 Heidelberg, Germany}
\email[show]{kalda@mpia.de}  

\author[0000-0001-5417-2260]{Gregory M. Green}
\affiliation{Max-Planck-Institut f\"{u}r Astronomie \\
K\"{o}nigstuhl 17 \\
D-69117 Heidelberg, Germany}
\email{green@mpia.de}

\begin{abstract}
The gravitational potential of the Milky Way encodes information about the distribution of all matter -- including dark matter -- throughout the Galaxy. \textit{Gaia} data release 3 has revealed a complex structure that necessitates flexible models of the Galactic gravitational potential.
We make use of a sample of 5.6 million upper-main-sequence stars to map the full 3D gravitational potential in a one-kiloparsec radius from the Sun using a data-driven approach called ``Deep Potential''. This method makes minimal assumptions about the dynamics of the Galaxy -- that the stars are a collisionless system that is statistically stationary in a rotating frame (with pattern speed to be determined). We model the distribution of stars in 6D phase space using a normalizing flow and the gravitational network using a neural network. We recover a local pattern speed of $\Omega_p = 28.2\pm0.1\mathrm{\,km/s/kpc}$, a local total matter density of $\rho=0.086\pm0.010\mathrm{\,M_\odot/pc^3}$ and local dark matter density of $\rho_\mathrm{DM}=0.007\pm0.011\mathrm{\,M_\odot/pc^3}$. The full 3D model exhibits spatial fluctuations, which may stem from the model architecture and non-stationarity in the Milky Way.
\end{abstract}


\section{Introduction} \label{sec:intro}

% Potential -> total matter density -> DM density. Axisymmetric models commonly used in the past (OPTIONAL: nice properties, such as integrability and action formalism). Gaia DR3 shows complex structure, necessitating more flexible models...

A central goal in galactic dynamics is to determine the spatial distribution of dark matter, which can only be traced through its gravitational effects. One gravitational probe of dark matter is the kinematics of stars in the Milky Way, which are guided by the gravitational potential, sourced in part by dark matter. Directly measuring stellar accelerations induced by the Galactic potential (typically in the order of \SI{1}{cm.s^{-1}.yr^{-1}}) is extremely challenging with current technology \citep{silverwood_easther_2019}, although some measurements exist for binary pulsars \citep{Moran24} and future probes are anticipated \citep{Chakrabarti20, Chakrabarti22}. Consequently, dynamical inferences primarily rely on an instantaneous snapshot of stellar positions and velocities provided by surveys like \Gaia. Without further assumptions about the dynamical state of the Galaxy, it is impossible to constrain the gravitational potential from a snapshot of stellar positions and velocities. A common assumption -- but not the only one -- is that the tracer population is statistically stationary in a chosen reference frame, typically the Galactocentric rest frame. Inserting this assumption into the collisionless Boltzmann equation (CBE) then generally yields an overconstrained system of equations, which one can use to solve for the gravitational potential. After recovering the total gravitational potential, which is sourced by the total matter density, one can obtain the dark matter density by subtracting a model for the baryonic component.

Historically, reconstructions of the gravitational potential have relied on simplifying assumptions about its form, such as axisymmetry. However, recent surveys like the \Gaia mission \citep{Gaia16, GaiaDR323} have highlighted the Milky Way's disk's true complexity, uncovering a wealth of non-axisymmetric and disequilibrium features, including the spiral arms, velocity waves, the Galactic warp, and the phase spiral. There is a complex interplay between the internal dynamics and external perturbations that drives the shape and growth of these features, meaning that it is often insufficient to look at one of these features in isolation.
To address this, we leverage the six-dimensional phase-space data of stars within one kiloparsec from the Sun and employ the ``Deep Potential'' method \citep{Green20,Green23}, which borrows methods from deep learning to model the distribution of matter and the local pattern speed within one kiloparsec from the Sun.

For a thorough overview of the literature on Milky Way dynamics, we refer to a recent review by \cite{Hunt25}. Traditional methods typically make assumptions about the form of the stellar distribution function (DF) or the gravitational potential, often motivated by a lack of data or its incompleteness. The most prominent approach is to use Jeans modeling, which takes velocity moments of the CBE at a given position to obtain a set of equations directly relating acceleration, and hence the gradients of the gravitational potential, to a set of observables. This is typically coupled with assumptions about spherical or rotational symmetries and isotropy in velocity dispersion \citep{BinneyTremaine2008}, though these assumptions can be partially relaxed \citep{Cappellari08}. Other distribution function based approaches include action-angle approaches \citep{Bovy13, Ting13}, orbit-superposition approach \citep{Schwarzschild79}, or the Made-to-Measure method \citep{Syer96}. One can additionally make use of chemical features as proxies for integrals of motion \citep{Price25} or for a completely different approach, utilize non-equilibrium features such as the Gaia phase spiral \citep{Widmark22, Guo24}.

The motivation of Deep Potential comes from the desire to make use of the full six-dimensional kinematic \Gaia data with minimal structural priors. This is achieved by fitting both a non-parametric DF, represented by a normalizing flow (NF), and a non-parametric gravitational potential, represented by a neural network (NN). 
This is coupled with a robust model for the \Gaia selection function in order to account for the incompleteness of the dataset. In contrast with typical approaches for modeling the selection function, we employ a spatially smooth model instead, as it would otherwise produce a non-continuous distribution function, which is difficult to model with normalizing flows. Here we summarize the assumptions of Deep Potential:

\begin{enumerate}
    \item The motions of stars are guided by a gravitational potential $\Phi(\vec x)$.
    \item We observe the phase-space coordinates of stars (the ``kinematic tracers'') that are statistically stationary in a rotating frame around the Galactic Center.
    \item The overall matter density is non-negative everywhere: $\rho(\vec x) \geq0$. We can express this via gravitational potential using the Poisson equation as $\rho(\vec x) = \laplacian \Phi/(4\pi G) \geq 0$.
\end{enumerate}

Our previous work on this method outlined its theoretical motivations and demonstrated its effectiveness on simpler toy models with observational errors and non-stationarities \citep{Green20,Green23}. \cite{Kalda24} demonstrated the effectiveness of the method on an $N-$body simulation of a barred galaxy and the recovery of its acceleration and density profiles along the global rotation speed.

Similar methods to Deep Potential, using normalizing flows to represent the stellar distribution function and then determining the gravitational potential by assuming stationarity, have been developed by \citet{An2021}, \citet{Naik2022GalacticAcc}, and \citet{Buckley2022GalacticDM,Buckley23}. \citet{Lim2023} recently applied normalizing-flow-based modeling to a sample of RGB stars within \SI{4}{kpc} from the Sun in order to estimate the local dark matter density and applied a different approach for accounting for the incompleteness of the observed distribution function by making use of the CBE being overconstrained and assuming the selection function to only depend on spatial coordinates. We also highlight that using dynamical mass measurements to study non-axisymmetric features has seen recent success with the usage of vertical Jeans modeling to detect spirals arms in the form of $\sim \SI{20}{\%}$ overdensities and by partitioning the Galactic disk plane into a $\SI{100}{pc}$ grid \citep{Widmark24}. 

In this paper, we describe the methodology of Deep Potential including accounting for the selection function (Section~\ref{sec:method}), describe the data selection (Section~\ref{sec:data_selection}), discuss the results in the Milky Way (Section~\ref{sec:results}), and future prospects (Section~\ref{sec:conclusion}).

\section{Method}\label{sec:method}
This work makes use of the Deep Potential method, which was first explained in \citet{Green20}, \citet{Green23}, and further developed to incorporate concurrent fitting of both the gravitational potential and the rotating frame in which the system appears most stationary in \citet{Kalda24}. The main methodological improvements in this work come in the form of aggregating multiple trained models to improve the accuracy of the result and in the uncertainty estimation pipeline. Similar to \cite{Putney2024}, we opt to train an ensemble of models by bootstrapping and resampling the errors in the kinematic data to estimate statistical uncertainties. In the following, we briefly review the key components of the Deep Potential method and then outline how we calculate error estimates.

\subsection{Overview of Deep Potential}\label{sec:overview deep potential}
The first assumption of Deep Potential is that stars orbit in a background gravitational potential, $\Phi (\vec x)$. The density of an ensemble of stars in six-dimensional phase space (position $\vec x$ and velocity $\vec v$) is referred to as the distribution function, $f(\vec x, \vec v)$. The evolution of the distribution function is described by the Collisionless Boltzmann Equation:
\begin{equation}
    \dv{f}{t} = \pdv{f}{t} + \sum_{i} \left(v_i\pdv{f}{x_i} - \pdv{\Phi}{x_i}\pdv{f}{v_i}\right) = 0.
\end{equation}

Our second assumption is that the distribution function appears stationary in a frame that is rotating with angular speed $\vec \Omega$ around an axis passing through a point $\vec x_0$ in space. This is akin to assuming that the distribution of stars rotates like a solid body around a central axis of the system. We additionally allow the frame in which stationarity holds to move with constant velocity $\vec v_0$ relative to the inertial measurement frame. In the Milky Way, where we obtain measurements in a Sun-centric frame from \Gaia, $\vec x_0$ and $\vec v_0$ represent the location and velocity of the Galactic Center relative to the Solar System, respectively. $\vec \Omega$ represents the best-fit pattern speed in the volume of interest and is expected to be a combination of the pattern speeds of spiral arms, moving groups, central bar, or the effects of other non-axisymmetric or disequilibrium features in the solar neighborhood. In general, the parameters describing the stationarity frame can either be fixed or determined concurrently with the gravitational potential. We opt to fix both $\vec x_0$ and $\vec v_0$ as those have been observationally measured to a very high degree of accuracy \citep{Gravity19}. We further fix the direction of $\vec \Omega$ to point towards the $z$-axis with a variable magnitude $|\vec\Omega|$ (see section \ref{sec:GDR3} for further discussion on the coordinate system). We thus have one free and six fixed variables describing the frame of best stationarity.

In the following, we denote the partial derivative of the distribution function with respect to time \textit{in the rotating frame} as $(\pdv*{f}{t})_\Omega$. The stationarity condition states that this partial derivative should equal zero. By translating the rotating-frame partial derivative to partial derivatives in the inertial measurement frame, the generalized stationarity condition in terms of inertial measurement-frame quantities is given by
\begin{align}
    \label{eqn:generalized-stationarity}
    \left(\pdv{f}{t}\right)_\Omega &=
      \pdv{f}{t}
      + \sum_{i}\left( u_i(\vec x) \pdv{f}{x_i}
      + w_i(\vec v)\pdv{f}{v_i}\right)
    = 0,\\
    \vec u(\vec x) &= \vec\Omega\times (\vec x - \vec x_0) + \vec v_0,\\
    \vec w(\vec v) &= \vec\Omega\times(\vec v - \vec v_0).
\end{align}
The terms $\vec u(\vec x)$ and $\vec w(\vec v)$ roughly correspond to the velocity change caused by moving into the rotating frame, and the fictitious force created by doing so, respectively. For a full derivation of this transformation, see \citet{Kalda24}.
Combining this with the Collisionless Boltzmann Equation, we arrive at
\begin{align}
    \label{eq:CBE+stat}
    \left(\pdv{f}{t}\right)_\Omega
    = \sum_i \left[
      (u_i - v_i)\pdv{f}{x_i}
      + \left(\pdv{\Phi}{x_i} + w_i\right)\pdv{f}{v_i}
    \right]
    = 0.
\end{align}
This results in an overconstrained equation for the gravitational potential, where a three-dimensional field has to satisfy CBE+stationarity in a six-dimensional space. If the DF in question was truly stationary, the gravitational potential could be uniquely determined by solving Eq.~\eqref{eq:CBE+stat}. However, realistic physical systems are never perfectly stationary. Consequently, a potential that perfectly satisfies Eq.~\eqref{eq:CBE+stat} might not exist (See \citealt{An2021} and \citealt{Green23} for discussion). In general, therefore, Deep Potential recovers the potential which minimizes some measure (to be discussed below) of the total non-stationarity in the system. We also emphasize that the assumption that the system is stationary at this moment in time does not require the gravitational potential to be time-independent in a rotating frame; rather, we are fitting the value of the gravitational potential at this moment in time and make no inferences on its time evolution. Additionally, we do not assume that the gravitational potential is sourced by the observed stellar population alone. Accordingly, we do not impose the condition
\begin{equation}
    \laplacian\Phi(\vec x) = 4\pi G \int f(\vec x, \vec v)\dd[3]{\vec v}.
\end{equation}
Additional mass components, such as dark matter, gas, and stellar populations not contained in our dataset, contribute to the total mass that we recover.

\subsection{Modeling the distribution function}

In practice, when we observe stellar populations, we obtain a discrete sample of points in phase space, rather than a smooth distribution function $f(\vec x,\vec v)$. To model the distribution function, we use normalizing flows, which produce a continuous, differentiable, and sampleable representation of the DF. Normalizing flows are a class of algorithms used for density estimation in unsupervised machine learning (for a review, see \citealt{Kobyzev2019}). A normalizing flow works by learning a set of invertible coordinate transformations that turn a simple distribution, usually a normal distribution, into a more complex one that matches the observed data. The various architectures used for normalizing flows are in a constant state of development, each with its own strengths and drawbacks. Continuing from previous work, we use FFJORD \citep{Grathwohl18} normalizing flows, though the particular choice of method is not critical for the method of this work. In general, the complexity of the distributions that the normalizing flows can capture is limited by the number of parameters describing the coordinate transformations and the specific architecture at hand. Key drawbacks of normalizing flows include their assumption of continuously distributed training data. Additionally, like many machine learning methods, they require careful hyperparameter tuning to avoid over- or underfitting and can be computationally intensive, especially with complex datasets.

For $n$ stars with positions $\vec x_i$, velocities $\vec v_i$, and weights $w_i$ (chosen to correct for variations in the observational selection function), we train a normalizing flow $f(\vec x, \vec v)$ using stochastic gradient descent to maximize the log-likelihood
\begin{align}
    L_f = \sum_{i=1}^{n} w_i \ln f\left(\vec x_i, \vec v_i\right). 
\end{align}
The loss is supplemented with Jacobian and kinetic regularization \citep{Finlay2020}, as detailed in Section~\ref{sec:implementation}. The final estimates for $f$ and its gradients $\pdv*{f}{\vec x}$ and $\pdv*{f}{\vec v}$ are obtained by averaging over 16 flows trained on the same data with different random seeds. We find that using multiple flows and taking the average yields better results than training a single, more complex flow. Additionally, training multiple flows allows us to estimate the uncertainty in the normalizing flow reconstruction (see section \ref{sec:uncertainty}).

\subsection{Modeling the gravitational potential}

After learning the distribution function, we find the gravitational potential $\Phi(\vec x)$ and pattern speed that best satisfies the CBE and generalized stationarity assumption given in Eq.~\eqref{eq:CBE+stat}. To parameterize the gravitational potential, we use a feed-forward neural network that takes as input a three-dimensional vector $\vec x$ and outputs a scalar, $\Phi$.
We simultaneously train the parameters of the potential and pattern speed $|\vec\Omega|$ to minimize
\begin{equation}
    \label{eq:grav_loss_int}
    L_\Phi = \int
      \mathcal L \left[
        \left(\pdv{f(\vec x,\vec v)}{t}\right)_\Omega
        \!\!\! , \,
        \laplacian \Phi(\vec x)
      \right]
      f(\vec x, \vec v)
      \dd[3]{\vec x}\dd[3]{\vec v},
\end{equation}
where $\mathcal L$ is the differential contribution to the loss of an individual point in phase space, given by
\begin{equation}
    \label{eq:grav_loss_individual}
    \mathcal L =
      \mathrm{arcsinh}\left[
        \alpha\left|\left(\pdv{f}{t}\right)_\Omega\right|
      \right]
    + \lambda~\mathrm{arcsinh}\left(
        \beta \max\left\{-\laplacian\Phi, 0\right\}
      \right).
\end{equation}
The first term penalizes non-stationarity in a frame rotating with angular velocity $\Omega_z\,\hat{z}$ about a point (which we assume to be the Galactic Center) with position and velocity offsets $\vec x_0$ and $\vec v_0$ from the Solar System. The second term penalizes negative mass densities. We first take the absolute value of $(\pdv*{f}{t})_\Omega$, in order to penalize positive and negative changes in the phase-space density equally. The inverse hyperbolic sine function down-weights large values, while the constant $\alpha$ sets the level of non-stationarity at which our penalty transitions from being approximately linear to being approximately logarithmic. The loss is supplemented with $\ell_2$ regularization on the neural-network weights that describe $\Phi(\vec x)$. The integral in Eq.~\eqref{eq:grav_loss_int} is computationally expensive to evaluate directly, but can be approximated by averaging $\mathcal{L}$ over $m$ samples drawn from the distribution function, where $m$ is a sufficiently large number:
\begin{equation}
    \label{eq:grav_loss_sum}
    L_\Phi \approx
      \frac{n}{m} \sum_{i=1}^{m}
        \mathcal{L} \left[
          \left(\pdv{f(\vec x_i,\vec v_i)}{t}\right)_\Omega
          \!\!\! , \,
          \laplacian \Phi(\vec x_i)
        \right].
\end{equation}
The constant $n/m$ comes from the normalization of the distribution function and can be omitted when implementing the loss function. Similarly to the distribution function, the final values for the gravitational potential are obtained by averaging over 16 different neural networks that have been trained with different random seeds on the same set of samples, drawn from the best-fit distribution function.

\subsection{Implementation}
\label{sec:implementation}

We implement Deep Potential in TensorFlow~2 \citep{tensorflow2015-whitepaper} and use TensorFlow Distributions \citep{Dillon17}. All of our code and trained models are publicly available under a permissive license that allows reuse and modification with attribution, both in archived form at \url{https://doi.org/10.5281/zenodo.16422931}
and in active development at \url{https://github.com/gregreen/deep-potential}.

To represent the distribution function, we train 16 different normalizing flows (and subsequently compute the median values among them), each with a chain of three FFJORD normalizing flows, each with 12 densely connected hidden layers, of 192 neurons and a $\tanh$ activation function. For our base distribution, we use a multivariate Gaussian distribution with mean and variance along each dimension set to match the training dataset. During training, we impose Jacobian and kinetic regularization with strengths \num{1e-5} \citep{Finlay2020}, which penalizes overly complex flow models and tends to reduce training time. We train our flows using the rectified Adam optimizer \citep{Liu2019}, with a batch size of $2^{13}$ (8096). We find that this relatively large batch size leads to faster convergence (in wall time) than more typical, smaller batch sizes. We begin the training with a ``warm-up'' phase that lasts approximately 3\% of an epoch, in which the learning rate linearly increases from 0 to \num{0.001}. Thereafter, we use a constant learning rate. We decrease the learning rate by a factor of two whenever the training loss fails to decrease below its previous minimum for 1024 consecutive steps (this period is termed the ``patience''). We terminate the training when the learning rate drops below \num{5e-6}.

After training our normalizing flow, we draw $m = 2^{22}$ ($\sim 4$ million) phase-space coordinates equally from the 16 normalizing flows within the volume outlined in \ref{sec:mask} and calculate the median gradients $\pdv*{f}{\vec x}$ and $\pdv*{f}{\vec v}$ at each point (using auto-differentiation), for use in learning the gravitational potential.

We represent the gravitational potential with the median of 16 feed-forward neural networks,\footnote{When taking derivatives of the potential (\textit{e.g.}, to calculate the acceleration field or the matter density), we always take the derivative of each neural network first, and then take the median of the derivatives. This avoids introducing discontinuities into the derivatives.} each with four densely connected hidden layers, each with 512 neurons and a $\tanh$ activation function. The network takes a three-dimensional input (the position $\vec x$ in space), and produces a scalar output (the potential). No activation function is applied to the final scalar output. We add in an $\ell_2$ loss on the potential network weights with strength $0.001/\mathtt{n\_weights}$, where \texttt{n\_weights} is the total number of weights in the network. The $\ell_2$ regularization modifies the log-loss of the neural network by adding an extra term, $\ell_2/\mathtt{n\_weights}\sum_{i}^{\mathtt{n\_weights}}w_i^2$, where $w_i$ is the value of the $i$-th weight in the network. This serves to penalize large values in the network and control over-fitting. We train the network using the rectified Adam optimizer, with batches of $2^{15}$ (\num{32768}) phase-space coordinates. We use a similar learning-rate scheme as before, with a warm-up phase lasting one epoch, an initial learning rate of \num{0.001}, a patience of 2048 steps, and a final learning rate of \num{1e-6}. In the potential loss function (Eq.~\ref{eq:grav_loss_individual}), we set $\alpha=\num{1e5}$, $\beta=1$, and $\lambda=1$, which affect the penalties on non-stationarity and negative gravitational mass densities.

When fitting both the distribution function and the gravitational potential, we reserve \SI{25}{\%} of our input data as a validation set. After each training step, we calculate the loss on a batch of validation data, in order to identify possible overfitting to the training data. Such overfitting would manifest itself as a significantly lower training loss than validation loss. In the experiments in this paper, no significant overfitting is observed -- the difference in the likelihoods of the training and validation sets is typically less than 1\%.

\subsection{Uncertainty estimation}
\label{sec:uncertainty}

We aim to quantify the following sources of error:
\begin{itemize}
    \setlength\itemsep{0pt}
    \setlength\parskip{0pt}
    \item Fluctuations that stem from the convergence of the trained models (by training multiple models and observing how they vary).
    \item Errors in the measured positions and velocities of the stars used in the training data (by resampling the positions and velocities of the stars using their reported uncertainties).
    \item Shot noise from the finite size of the dataset (by resampling the dataset with replacement using bootstrap).
    \item Errors in the literature values of $\vec x_0$, and $\vec v_0$ (by resampling their values before training the gravitational potential using their reported uncertainties).
\end{itemize}
We provide conservative estimates for all of these error sources and add them together in quadrature, where appropriate. However, we do not account for all sources of error, including normalizing flows systematically having difficulties capturing sharp features, and the stationarity condition not holding in real systems, among others. Hence, the quoted errors should be taken as a lower bound on the actual errors. For a full overview of the uncertainty modeling pipeline, see Appendix~\ref{appendix:uncertainties}. In particular, Figure~\ref{fig:dpot_diagram} gives a visual overview of the modeling pipeline, and Figure~\ref{fig:scalar_values} shows the relative contributions of the different sources of error to the final model outputs. These should be interpreted with care, though, as some of the contributions are strongly correlated with each other.

\begin{table}[]
\centering
\begin{tabular}{r|p{60mm}}
Quantity    & Description                                                          \\\hline
$r$         & Distance from the Sun                                                \\
$R$         & Cylindrical radius in the Galactocentric coordinate system           \\
$\phi$      & Azimuthal angle in the right-handed Galactocentric coordinate system \\
$(v_R, v_\Phi)$ & Velocity components along cylindrical radius and azimuthal angle in the right-handed Galactocentric coordinate system                                      \\
$z$         & Height above the Sun in the Galactic coordinate system               \\
$(x, y)$    & Cartesian coordinates in the Galactic coordinate system              \\
$(\ell, b)$ & Galactic longitude and latitude                                     \end{tabular}
\caption{
Definitions of the various coordinates used in this paper.}
\label{tab:coordinates}
\end{table}

\section{Data selection} \label{sec:data_selection}

We select a high-quality subset of stars with full six-dimensional position and velocity measurements from the ESA \Gaia space telescope, within 1\,kpc of the Sun.

\subsection{Gaia DR3 spatial and quality cuts}
\label{sec:GDR3}

\Gaia has measured astrometry for over a billion sources in the Milky Way. Of those, \num{33812183} stars have full six-dimensional phase-space information. We first apply the \Gaia parallax zero-point correction based on \citet{Lindegren21} and then apply a set of quality cuts, outlined in table~\ref{tab:cuts}, including a cut on fidelity derived from \citet{Rybizki23}. We are left with \num{26519644} sources, from which we remove prominent open clusters (section \ref{sec:OC}).

We then select a volume of $r < \SI{1000}{pc}$ from the Sun. This distance range is sufficient to adequately capture the vertical profile of the Milky Way disk, while not extending too far into the dust clouds, which are more computationally expensive to model. When training normalizing flows for the distribution function, we model the effect of the selection function (section \ref{sec:sel}) by upweighing stars according to the inverse of their selection probability (section \ref{sec:final_dataset}), and further pad our selected volume by adding stars outside the volume, in a manner that smoothly tapers off the density of selected stars. This facilitates the training process, similarly to \cite{Kalda24} (section \ref{sec:final_dataset}). Finally, before training the gravitational potential, we mask out volumes that we deem unreliable due to a lack of data, usually due to crowding or dust clouds (section \ref{sec:mask}).

We use a right-handed Galactic coordinate system \citep{Blaauw60}, which is centered on the Solar System barycenter both in position and velocity, with the $x$-axis pointing towards the Galactic Center and the $z$-axis towards the North Galactic Pole. The axis of rotation then passes through $\vec x_0 = (R_0, 0, 0)$, the rotation vector is $\vec\Omega = (0, 0, \Omega_z)$, and $\vec v_0$ is interpreted as the velocity of the Galactic Center with respect to the Solar System. We fix the distance to the Galactic Center as $R_0 = \SI{8.277\pm 0.04}{kpc}$, and $\vec v_0 = (\num{9.3\pm 1.3},\num{251.5\pm 1.0},\num{8.59\pm 0.28})\,\si{km/s}$ \citet{Gravity22}. We note that with this choice of coordinates, $\Omega_z$ is negative due to the Milky Way rotating clockwise around the North Galactic Pole. We, therefore, report the absolute value of the rotation speed in the subsequent analysis and denote it as $\Omega_p \equiv \left| \Omega_z \right|$, where $p$ stands for the pattern speed. The notation we use for various coordinate systems is outlined in Table~\ref{tab:coordinates}. 

\begin{table}
\centering
\begin{tabular}{c}
    \Gaia DR3 cuts \\\hline
    has radial velocity  \\
    has \texttt{G-Rp} \\
    $\texttt{phot\_g\_mean\_mag} < 15$\\
    $\texttt{fidelity\_v2} > 0.5$ \\
    $\texttt{parallax} > \SI{0}{mas}$\\
    $\texttt{parallax\_error} < \SI{0.05}{mas}$\\
    $\texttt{ra\_error}, \texttt{dec\_error} < \SI{0.05}{mas}$\\
    $\texttt{pmdec\_error}, \texttt{pmra\_error} < \SI{1}{mas/yr}$\\
    $\texttt{error\_vr} < \SI{10}{km/s}$\\
\end{tabular}
\caption{
Cuts performed on the \Gaia DR3 catalog. \texttt{fidelity\_v2} is the ``astrometric fidelity'' rating from \citep{Rybizki23}, which classifies astrometric solutions as reliable or unreliable. Out of \num{33812183} stars with radial velocity measurements, \num{26519644} pass our quality cuts.}
\label{tab:cuts}
\end{table}

\subsection{Removing open clusters}\label{sec:OC}

Open clusters present themselves as significant overdensities compared to the overall background distribution of stars. We remove known open cluster members for several reasons: their high phase-space densities can introduce systematics when fitting the smooth DF, their inherent non-stationarity violates a key model assumption, and they constitute a small fraction of the total stellar mass ($\sim \SI{0.2}{\%}$). We disentangle the open clusters from the background by selecting the list of candidate stars for each open cluster in \citet{Hunt23}, using their suggested cuts on fidelity. We only consider open clusters within \SI{4}{kpc} from the Sun, and that have at least 20 members within our dataset. We additionally filter out the Orion Nebula, as it historically has not been included in open cluster catalogs despite fulfilling the criteria. We do this by applying a Gaussian mixture model using the expectation-maximization algorithm and modeling the Orion Nebula using a Gaussian profile in sky position and proper motion located near $\ell=\ang{205}, b=\ang{-19.5}$, and the background of stars as a Gaussian profile in proper motion alone (and uniform density in sky position). In the end, we filter out \num{39233} stars from 694 open clusters for which the membership probability exceeds \num{0.5}, leaving \num{26480411} stars.

\begin{figure*}
    \includegraphics[width=\linewidth]{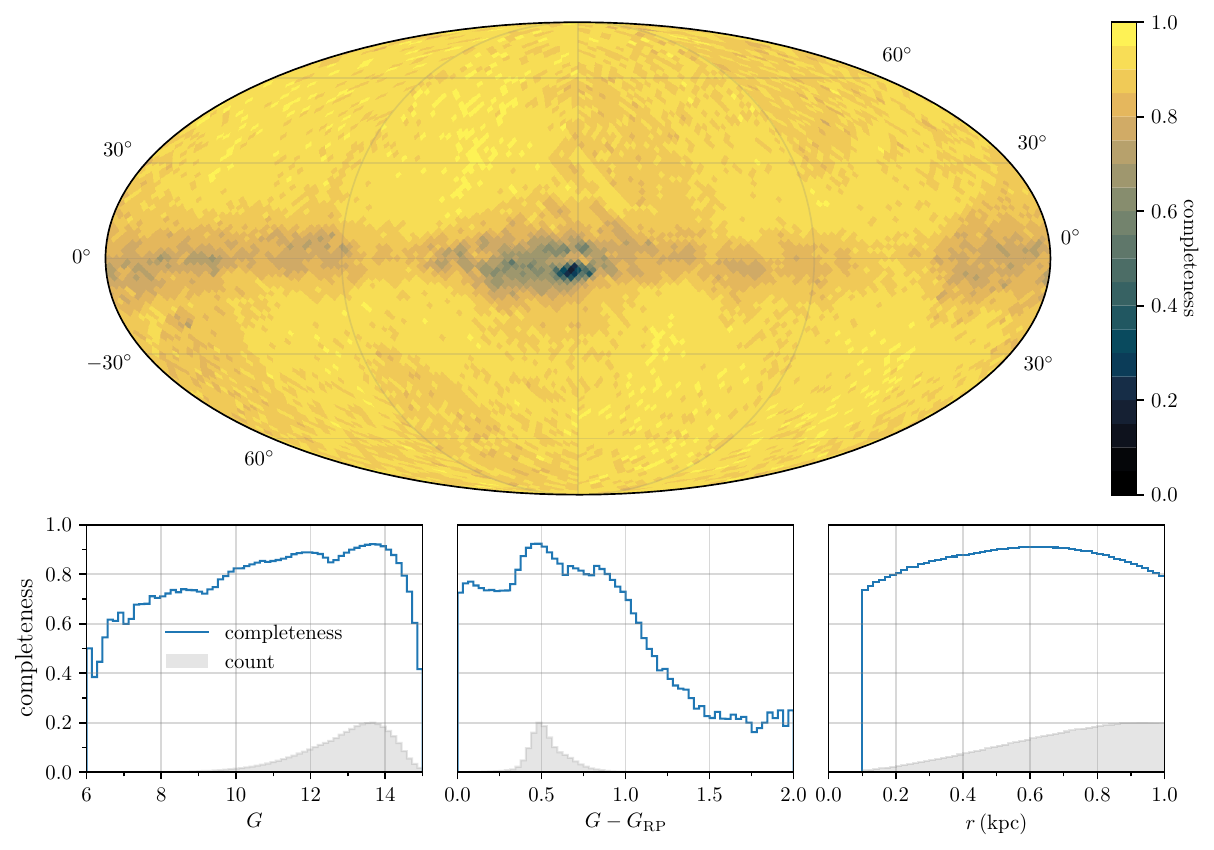}
    \caption{Mean completeness of the \num{5617061} sources in the \SI{1000}{pc} dataset with full 6D phase space kinematics in \Gaia DR3 that pass the quality cuts specified in section \ref{sec:GDR3} as a function of sky-position (top subplot), apparent magnitude $G$, color $G-G_\mathrm{RP}$ and extinction $E$ from \citet{Edenhofer23} of the source (bottom row). Each panel in the bottom row additionally includes a histogram of the sources in the dataset (light gray). The completeness represents the probability that a source with given physical parameters is in the dataset and is estimated by calculating the ratio of the number of sources in the dataset to the number of sources within the parent \Gaia DR3 catalog with $G < 15$. We observe the dependence of completeness on crowding (especially at Baade's window), apparent magnitude, color, and a weak dependence on the scanning pattern of \Gaia.}
    \label{fig:completeness}
\end{figure*}

\begin{figure*}
    \centering
    \includegraphics[width=\linewidth]{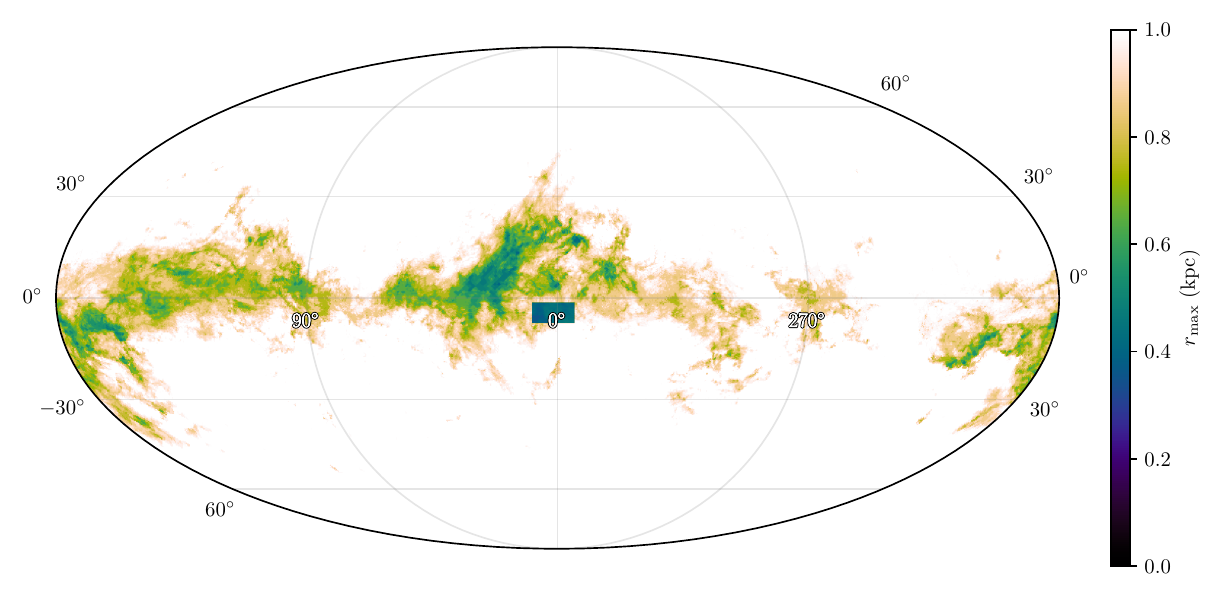}
    \caption{Maximum distance $r_\mathrm{max}$ within which $>95\%$ of the \SI{1}{kpc} \Gaia dataset stars would pass the apparent magnitude limit, $G<\num{14.8}$. The crowded Baade's window near the Galactic Center is handled separately with a brightness limit of $G < 13$.}
    \label{fig:max_dist}
\end{figure*}

\subsection{Modeling the \Gaia selection function}
\label{sec:sel}

The Deep Potential method, as sketched out above, requires us to observe a population of phase-mixed kinetic tracers (stars) within some subvolume of the Galaxy. However, due to effects such as dust obscuration and varying observational depth, we do not observe the full population of stars that we seek to target. If selection effects introduce unphysical gradients into the distribution function of observed stars, then our recovered gravitational potential will be affected. We, therefore, model the selection function and correct our dataset to obtain an estimate of the underlying, true distribution function of our target stellar population.

The selection function is defined as the probability of a source with given observational parameters being included in the dataset. It is a function of the observational properties of the source, such as its apparent magnitude, color, and crowding, but also effects arising from the processing pipeline and the scanning pattern of the \Gaia satellite. As we are dealing with a subset of the GDR3 catalog, we can express the selection function as
\begin{equation}
    S(\vec q) = S(\vec q\: |\, \vec q\text{ in GDR3}) \cdot S_\mathrm{GDR3}(\vec q),
\end{equation}
where $S_\mathrm{GDR3}(\vec q)$ is the probability that a source with attributes $\vec q = \{\text{sky-position}, G, \text{color}, \ldots\}$ is in the \Gaia catalog and $S(\vec q\: |\, \vec q\text{ in GDR3})$ is the probability that the source is in the subsample, given that it is in the \Gaia parent catalog. In the following, we assume that $S_\mathrm{GDR3}(\vec q) = 1$ within our subset of interest, as GDR3 is close to fully complete within our range of apparent magnitudes \citep{Cantat23}.

There exist various approaches to modeling the selection function of the Milky Way, such as defining bins of sky-position, color, and apparent magnitude, and taking the ratio of sources in the sub-catalog to the number of sources in the parent catalog \citep{Rybizki21, Ginard23} or using spherical needlets on the sky with weights derived from color and apparent magnitude \citep{Everall22}. While these are sufficient for many science cases, for Deep Potential, we require a spatially smooth model of the selection function. Otherwise, discontinuities in the selection function could introduce spurious gradients in the modeled stellar DF, potentially leading to unphysical densities. This can be especially important in areas with detailed small-scale structures, such as the vicinity of dust clouds.

We opt for the approach of modeling the selection function using a fully connected neural network $S_\theta (\vec q)$, which allows us to guarantee that our model of the selection function will be smooth. The neural network has parameters $\theta$ and a single sigmoid function as the final layer, guaranteeing that the output is between zero and one. We use 20 features as input parameters: three for describing the sky-position in terms of the components of a unit vector pointing towards the source, one for apparent magnitude in $G$-band, one for color $G-G_\mathrm{RP}$, one for extinction derived from \cite{Edenhofer23}, four for \Gaia features pertaining to the scanning pattern and ten for describing the correlations between the five primary astrometric solutions. We did not include additional features in order to avoid over-fitting and because the resulting models were already sufficiently spatially smooth. Notably, the choice of features implicitly captures crowding from sky-position.  We train the neural network on a dataset of stars from \Gaia DR3 with $\texttt{phot\_g\_mean\_mag} < 15, \texttt{fidelity\_v2} > 0.5$ and by minimizing binary cross-entropy
\begin{equation}
\theta = \mathop{\arg \min}\limits_{\theta}\sum_i \left[ p_i \ln S_\theta (\vec q_i) + (1 - p_i) \ln\left( 1 - S_\theta (\vec q_i)\right) \right].
\end{equation}
The sum is taken over all the training stars, $\vec q_i$ is the vector of features for source $i$. $p_i$ indicates whether a star is in our subset (value 1) or not (value 0).

Figure \ref{fig:completeness} shows the completeness in various projections. As can be seen, the selection function has dependencies on crowding, apparent magnitude, color, scanning pattern, and various other features. For brevity, we list the feature importance of the different input parameters in Appendix \ref{appendix:feature_importance}.

\begin{figure}
    \centering
    \includegraphics[width=\linewidth]{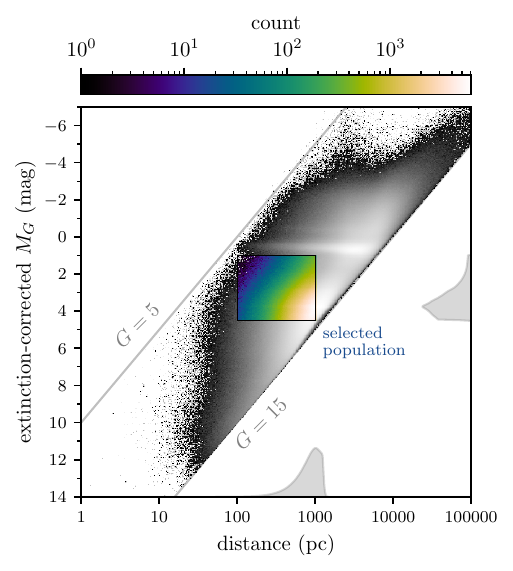}
    \caption{Histogram of extinction-corrected absolute magnitude versus distance from the Sun for the \num{5617061} sources in the final dataset (in color) compared to the stars excluded from the dataset that have full 6D phase-space measurements and pass our quality cuts (in gray). Along each axis, we additionally show 1D histograms of all sources with 6D phase-space measurements. The stars with 6D phase-space measurements are nearly uniformly complete between apparent magnitudes $5 < G < 15$. The final dataset is selected to have spatially uniform completeness for all sources within the chosen range of extinction-corrected absolute magnitudes, resulting in a box that extends from $\rin=\SI{100}{pc}$ to $\rout=\SI{1000}{pc}$ in the above diagram.}
    \label{fig:final_dataset_dist_M}
\end{figure}

\subsection{Constructing the final dataset}\label{sec:final_dataset}
Based on the dataset of stars that pass the quality cuts, have open clusters removed, and have an accompanying selection function $S_i$ for each source, we construct the final dataset. One of the assumptions of the method is that the ensemble of stars is statistically stationary. To this end, we could use external catalogs to make cuts on age or elemental abundances in order to select stars that are more phase-mixed and share a similar dynamic history. However, doing so would introduce additional selection effects, which are more difficult to model and lower the overall completeness of the data. We instead opt not to rely on external catalogs and apply cuts only on the extinction-corrected absolute magnitude $M_G$ in the $G$-band. We calculate the absolute magnitude using $M_G = G + 5\log_{10}\left(\varpi/\SI{100}{mas}\right) - A_G$ and extinction estimates in $G$-band \citep{Zhang2024}.

We apply spherical cuts such that the population forms a spherical shell extending from an inner radius of $\rin = \SI{100}{pc}$ to an outer radius of $\rout = \SI{1000}{pc}$ and has absolute magnitudes between $M_{G,1} = 1$ and $M_{G,2} = \num{4.5}$. The choice of values is motivated by avoiding extending too deep into dust clouds while maximizing the number of stars within the dataset and also respecting the brightness limits $G = 5$, $G = 15$ of \Gaia. This results in \num{5617061} sources in the final dataset.

Following \cite{Kalda24}, we include a `padding' of stars both inside and outside the boundaries of the dataset. This helps avoid sharp edges in the training data, which can be difficult for normalizing flows to model accurately.

\begin{figure*}
    \centering
    \includegraphics[width=0.95\linewidth]{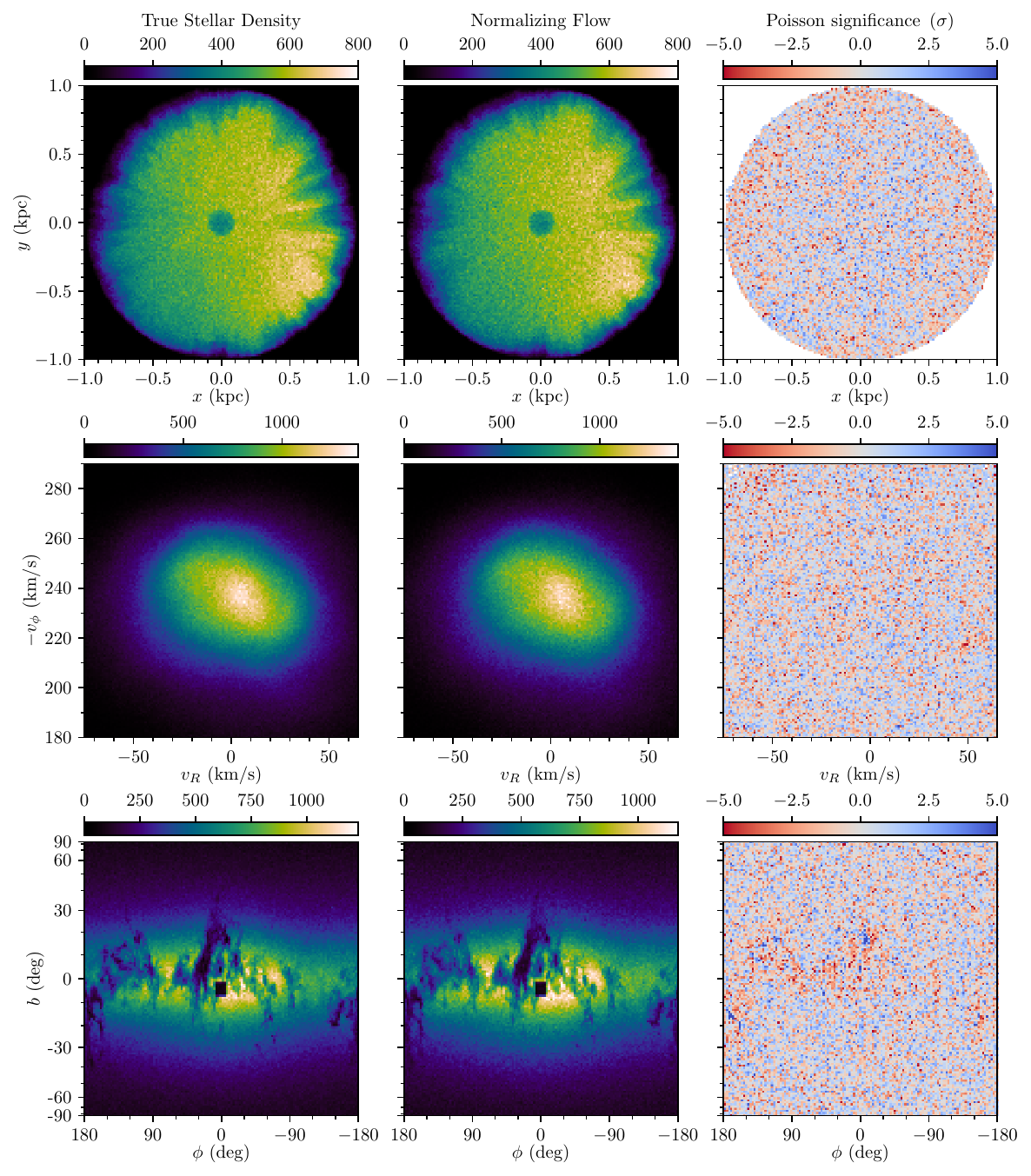}
    \caption{A demonstration of the performance of our normalizing flow model of the stellar phase-space distribution function in three different projections. We plot 2D histograms of the selected stars (left column) and of $2^{21}=2097152$ points sampled from the trained normalizing flow (middle column), as well as a comparison between the two (right column). The top row shows a face-on view of the \SI{1000}{pc} dataset, projected onto the $x-y$ plane. The middle row shows one velocity-space projection ($v_{\phi}$ vs. $v_R$), and the bottom row shows the sky projection (in a Lambert cylindrical equal-area projection: $\ell$ vs. $\sin b$). The density of the sampled normalizing flow points has been renormalized by an overall constant to match the density of the stars in the training set. For each bin, we define the Poisson significance as $(n_\mathrm{NF} - n_\mathrm{data})/\sqrt{n_\mathrm{data}}$, where $n_\mathrm{NF}$ is the renormalized number of samples in the bin drawn from the normalizing flow and $n_\mathrm{data}$ is the number of sources in the dataset in the same bin. As can be seen above, our normalizing flow captures nearly all prominent features in the stellar distribution, with the exception of some residuals seen in areas of high dust extinction in the sky projection. We thus obtain a smooth, differentiable representation of the Galactic stellar population.}
    \label{fig:df}
\end{figure*}

In order to account for the stars that are not present in the dataset due to the non-uniform selection function, we assign a weight of $w_i = S_\mathrm{target} / S_i(\vec q_i)$ to each star. We further clip the weights to be between $w_\mathrm{min}=0.1$ and $w_\mathrm{max} = 3$. The upper limit has a smooth roll-off and is set to avoid amplifying the Poisson noise too much. The lower limit serves to avoid having the stars with low weights dominating the dataset, especially within the padding. For stars with $w < w_\mathrm{min}$, we further reject them with probability $1 - w / w_\mathrm{min}$. For the target completeness, we fix it to $S_\mathrm{target}=1$ inside $\rin < r < \rout$ and let it vary spatially outside of the inner volume in order to conveniently account for the padding. We let $S_\mathrm{target} = 1$ when $r < \rin$ and $S(\vec r) = f(r-\rout, 0.1\rout)$ when $r > \rout$. Here, $f(x, \sigma)$ is the normal distribution with standard deviation $\sigma$. This effectively selects all sources inside $r<\rin$ with the same likelihood as inside the volume of validity, and selects stars outside of $\rout$ with a smooth roll-off following a normal distribution with a scalelength of $0.1\rout$. The normal distribution for the roll-off is motivated by the base distribution of a normalizing flow also being a normal distribution. The final dataset is visualized alongside the final model for the distribution function in Figure \ref{fig:df}.

\subsection{Masking out incomplete volumes}\label{sec:mask}
Regardless of how well we account for the selection function, there will be some volumes, especially towards the Galactic Center and disk, where the completeness will be too low to meaningfully recover the true distribution function. Before we train the gravitational potential, we mask out any such volume.

In order to construct the mask, we use a rough criterion that a star is ``fully recoverable'' if its apparent magnitude, after passing through dust, is $G < \num{14.8}$. This value is chosen such that a star, regardless of its extinction, would have a completeness of at least $\sim 1 / 3$. The mask is then defined by a maximum distance $r_\mathrm{max}(\ell, b)$ up to which at least \SI{95}{\%} of the stars that follow the same absolute magnitude distribution as the stars within our dataset are rendered ``fully recoverable.'' Extinction from dust clouds reduces the maximum distance. Crowding similarly reduces the maximum distance by reducing the limiting apparent magnitude.
We handle the extremely crowded ``Baade's window'' region ($\ell \in [\ang{-6}, \ang{9.2})$, $b\in [\ang{-7}, \ang{-1})$) with a more stringent cut of $G < 13$.

The final mask is visualized in Figure~\ref{fig:max_dist} and portrays the maximum distance of our mask as a function of sky position, $r_\mathrm{max}(\ell, b)$. Notably, this mask is only applied after training the distribution function on the full dataset with padding, before training the gravitational potential.

\begin{figure*}
    \centering
    \includegraphics[width=\linewidth]{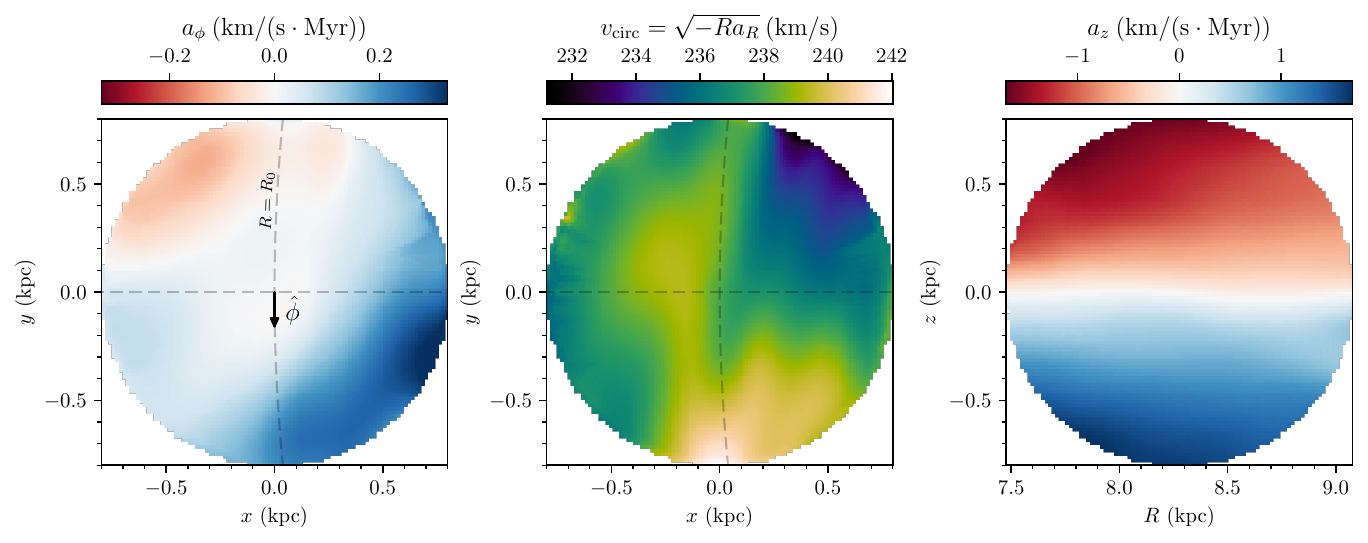}
    \caption{Two-dimensional projections of the model acceleration $\vec a = -\vec\nabla \Phi(\vec x)$ within $r < \SI{0.8}{kpc}$. The left panel shows the average azimuthal acceleration in the $x-y$ plane. The middle panel the average circular velocity in the $x-y$ plane, calculated as $\sqrt{-R a_R}$. The right panel shows the average vertical acceleration in the $R-z$ plane. The projections are obtained by averaging over the third spatial dimension. For the left and middle panels, we integrate over $\SI{-0.25}{kpc} < z < \SI{0.25}{kpc}$, while for the right panel, we integrate over $\SI{-0.25}{kpc} < y < \SI{0.25}{kpc}$ (holding $R$ constant).}
    \label{fig:all_acc_2d}
\end{figure*}

\begin{figure*}
    \centering
    \includegraphics[width=\linewidth]{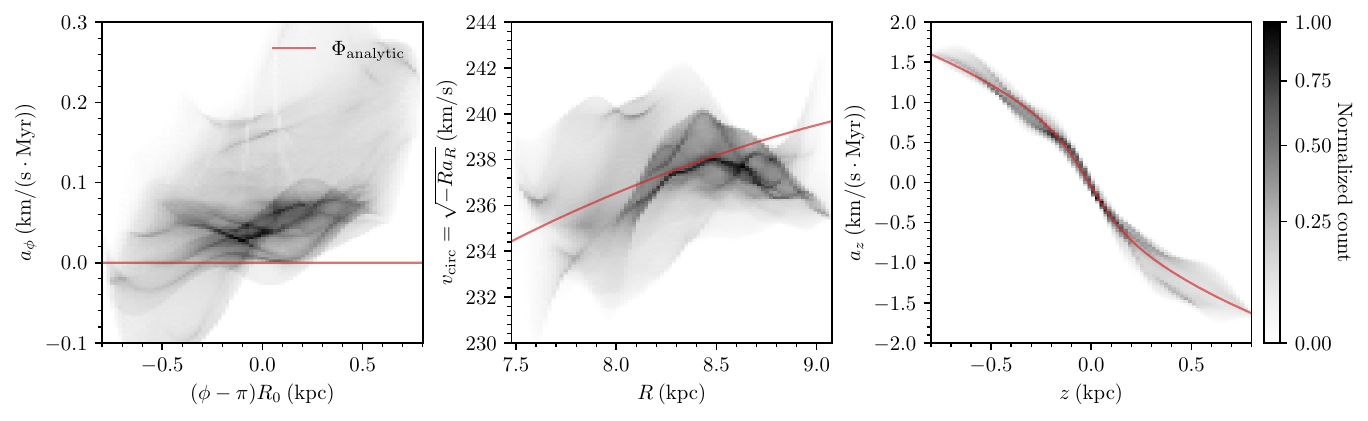}
    \caption{Comparisons of various one-dimensional projections of the accelerations $\vec a = -\vec\nabla \Phi(\vec x)$ for the full model (in gray) and a simple analytic model fit (in red). The gray background in each panel shows the amount of volume that has a given acceleration (or circular velocity), as a function of a single spatial coordinate.
    %The projections for the full model are obtained by marginalizing the accelerations within \SI{0.8}{kpc} from the Sun along the cylindrical radius $R$, tangential coordinate $(\phi - \pi)R_0$, and $z$.
    The left panel shows azimuthal acceleration vs. azimuth, the middle panel shows circular velocity $\sqrt{-Ra_R}$ vs. $R$, and the right panel shows vertical acceleration vs. height above the Sun. The analytic model is made up of three Miyamoto-Nagai disks and a Navarro-Frenk-White profile for the halo and is evaluated along the respective axes, passing through the Sun's location.}
    \label{fig:all_acc}
\end{figure*}

\begin{figure*}
    \centering
    \includegraphics[width=\linewidth]{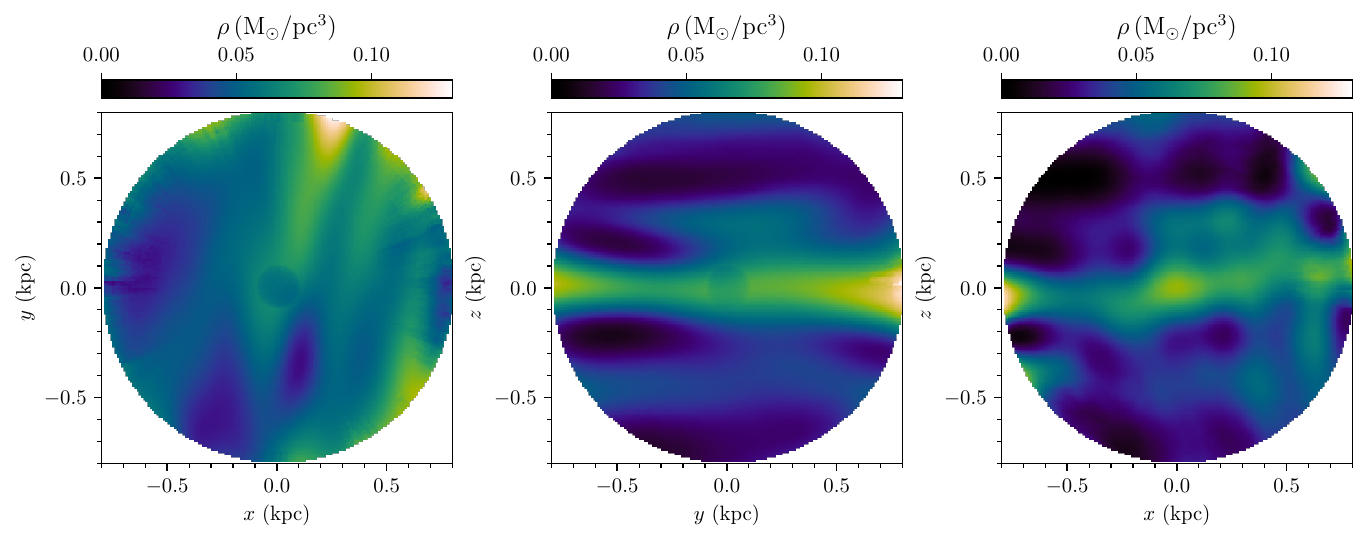}
    \caption{Two-dimensional projections of the model density $\rho = \nabla^2\Phi(\vec x) / (4\pi G)$ within $r < \SI{0.8}{kpc}$. The projections are in the $x-y$ plane (left panel), $y-z$ plane (middle panel), and $x-z$ plane (right panel), and are averaged over the respective third spatial Cartesian coordinate, between \SI{-0.25}{kpc} and \SI{0.25}{kpc}. The resulting projections show the disk of the Milky Way, but exhibit significant fluctuations, which may have a number of causes. Taking second derivatives of our fully nonparametric potential will amplify small, spurious fluctuations (compare with the smoother acceleration fields in Figure~\ref{fig:all_acc_2d}, based on the first derivatives of the potential). Real non-stationarities in the Milky Way, which violate our assumptions, may manifest as spurious density fluctuations in our model. Some of the fluctuations may also trace real variations in the underlying density field. Note that the inner 100~pc (where we have no tracer stars) is excluded from our calculations of the average density, causing visible, sharp density changes in the left two panels.}
    \label{fig:rho_2d}
\end{figure*}

\begin{figure}
    \centering
    \includegraphics[width=\linewidth]{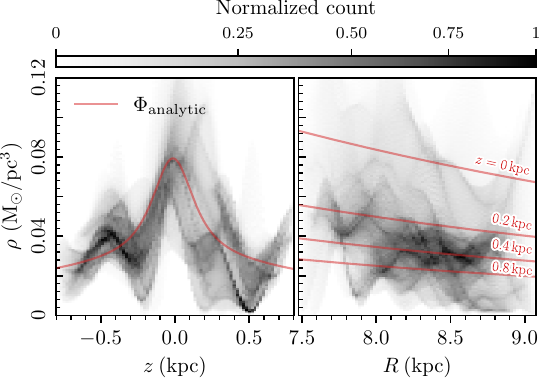}
    \caption{Vertical (left panel) and radial (right panel) density profiles for the full nonparametric model (gray) vs. a simple analytic model fit (red). The gray background in each panel shows the amount of volume that has a given density, as a function of a single spatial coordinate. The analytic model is made up of three Miyamoto-Nagai disks and a Navarro-Frenk-White profile, and is evaluated along $x=y=0$ in the left panel and $y=0$ with along varying levels of $z=\text{const}$ for the right panel. We observe significant fluctuations in the full nonparametric model density, which may arise in part from density estimates $\rho = \nabla^2 \Phi(\vec x)/(4\pi G)$ being sensitive to small fluctuations in the nonparametric potential model, non-stationarity in the Milky Way, or real density fluctuations in the Milky Way.}
    \label{fig:all_rho_z}
\end{figure}

\begin{figure}
    \centering
    \includegraphics[width=\linewidth]{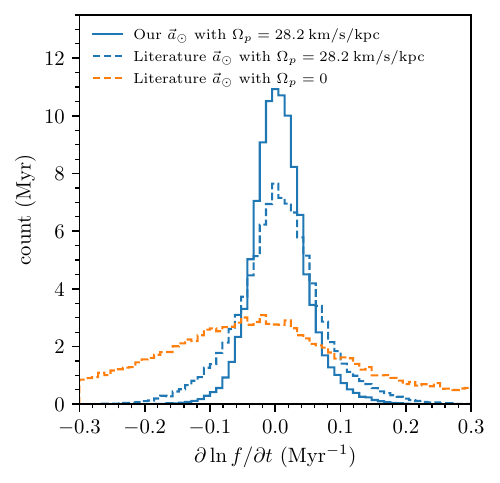}
    \caption{Comparison of the residual non-stationarity $\partial\ln f/\partial t$ near the solar neighborhood (averaged within \SI{200}{pc} from the Sun) for two models for the gravitational potential, one assuming the best-fit rotation speed $\Omega_p = \SI{28.2}{km/s/kpc}$ and another without rotation $\Omega_p=0$. The residue may represent inherent non-stationarity in the Milky Way, as well as imperfections in our models of the distribution function and gravitational potential. We observe that the characteristic timescale for changes within the phase-space density is in the order of \SI{20}{Myr} and \SI{5}{Myr} for the rotating and non-rotating cases, respectively.}
    \label{fig:hist1d_dlnf_dt}
\end{figure}

\begin{figure*}
    \centering
    \resizebox{\textwidth}{!}{\input{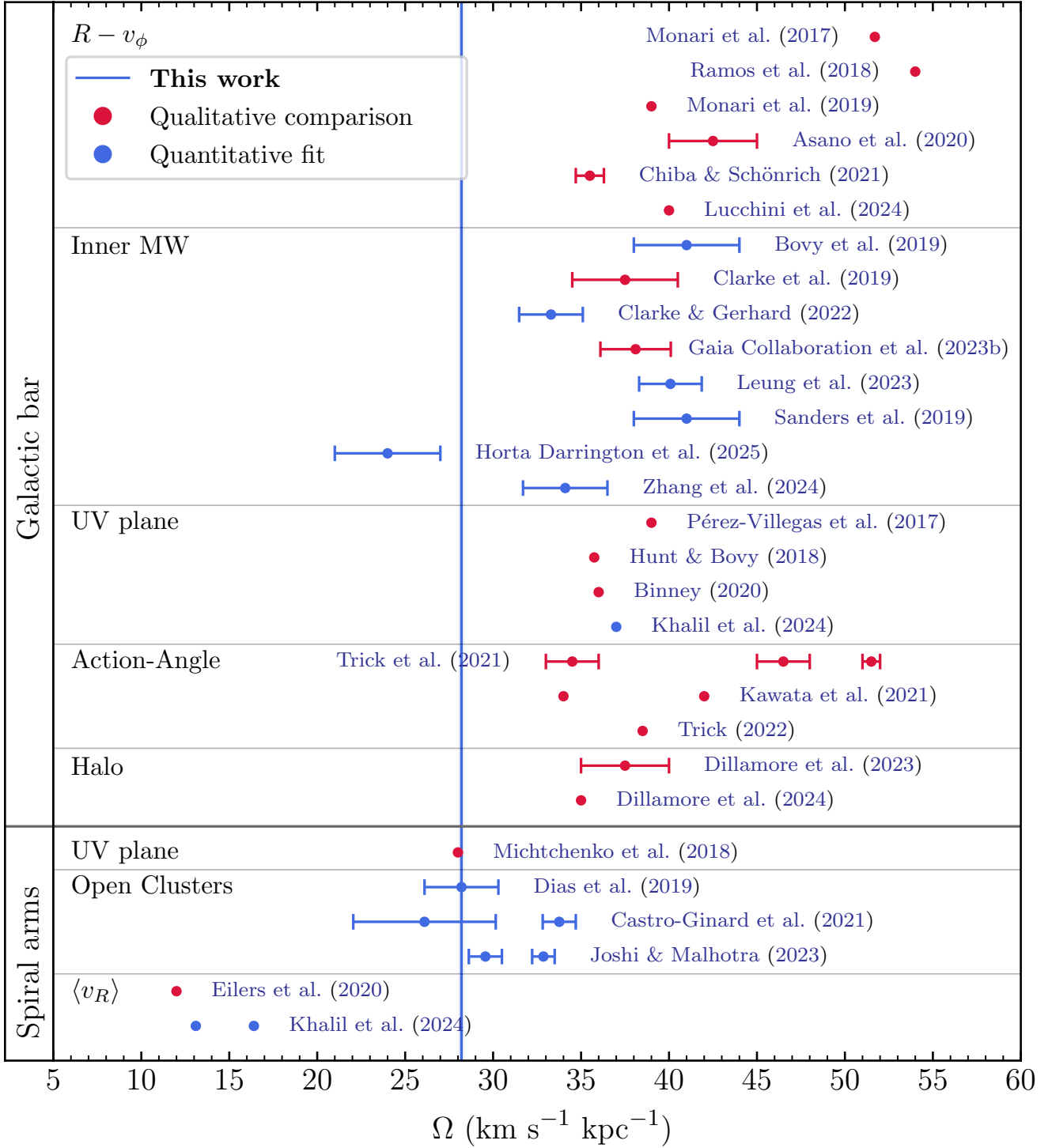}}
    \caption{Comparison of the value for the best-fit pattern speed (vertical blue line) within \SI{1}{kpc} of the Sun from this work and literature estimates (shown as dots) for the pattern speeds of the Galactic Bar and spiral arms in the Milky Way, adapted from Figure~10 of \citet{Hunt25}. Uncertainties are provided when available. Qualitative estimates are shown in blue and quantitative estimates in red. Some works provide either multiple pattern-speed models \citep{CastroGinard21,Joshi23,Khalil24} or multiple values for the respective feature \citep{Trick21,Kawata21}.}
    \label{fig:omega_literature}
\end{figure*}

\section{Results}
\label{sec:results}

\subsection{Model of the distribution function}

We obtain a normalizing-flow model of the six-dimensional distribution function, $f(\vec x, \vec v)$, within 1\,kpc of the Sun. There are various diagnostics for the performance of this distribution-function model. One way to test the performance of our distribution function model is to compare various two-dimensional projections of the modeled phase space density with the distribution of the stars in our training dataset (See Figure~\ref{fig:df}).

The normalizing flow captures the overall phase-space structure well, including known non-stationary features such as the phase-space spiral, moving groups, and velocity ripples. 
We observe no significant systematic differences between the density of the training data and the normalizing flow model, except in sky projections of regions of high extinction near the Galactic plane. These deviations arise from the complex morphology and sharp changes in the density of the training data, due to foreground extinction. This can be remedied at the expense of longer training times and more complex models, or potentially from a different choice of normalizing-flow architecture.
%Additionally, one could further facilitate training by artificially smoothing out volumes that are difficult to model.

Another diagnostic is to consider variation among our 16 different normalizing flows. By comparing $f, \pdv*{f}{\vec x},$ and $\pdv*{f}{\vec v}$ between flows, we can estimate the variance in these quantities introduced by the choice of random seed used for training. We find that the averaged, best-fit flow density has a 1$\sigma$ uncertainty of $\sim\SI{2.5}{\%}$, while the gradients have uncertainties of $\sim\SI{20}{\%}$. These uncertainties are relative to a hypothetical distribution function model created by averaging an infinite number of normalizing flows. There is no guarantee that an infinite number of NFs would converge towards the true underlying DF, though toy simulations of Plummer Sphere in \citet{Green20} did find good convergence towards the true DF. As discussed in subsequent subsections, the NF variation from the starting seed is one of the main contributors to the statistical errors in the final results.

\subsection{Modeling the gravitational potential}

After training the gravitational potential, we obtain a fully differentiable three-dimensional representation of the gravitational potential $\Phi(\vec x)$ near the Sun. At any specific point, we can calculate the acceleration $\vec a(\vec x) = -\vec\nabla\Phi(\vec x)$ and density $\rho(\vec x) = \nabla^2\Phi(\vec x)/(4\pi G)$ by computing the median value over the 16 models that were trained from the best-fit distribution function. We also perform simple four-component fits on the distribution function using three Miyamoto-Nagai disks and a Navarro–Frenk–White profile for the baryonic disk and dark matter halo, respectively. This serves as a comparison between the fully flexible potential model and a simpler axisymmetric model.

\subsubsection{Accelerations}

In Figure~\ref{fig:all_acc}, we plot one-dimensional profiles of acceleration vs position for three components, Galactocentric radius $R$, Galactic azimuth $\phi$, and height above the Sun $z$. In each panel, we show the results of our fully flexible model (in gray), as well as the results of the axisymmetric analytic potential (in red). Figure \ref{fig:all_acc_2d} shows the acceleration in two-dimensional projections, meant to highlight the non-axisymmetric structures. In all diagrams, the outer \SI{200}{pc} are omitted due to edge effects affecting the neural network and to avoid more heavily extincted volumes on the Galactic plane. We observe that the deviation between the simple axisymmetric model and the full model, i.e. $|\vec \nabla \Phi - \vec \nabla \Phi_\mathrm{analytic}| / |\vec \nabla \Phi|$, is within \SI{2}{\%} for the majority of the volume.
%This is, however, unsurprising, as the dominant forces -- the radial force towards the Galactic Center and the restoring force in the Galactic midplane -- are both axisymmetric.
Notably, the azimuthal acceleration $a_\phi$ is nonzero: the solar neighborhood is being pulled in a direction opposite to the Galactic rotation. This is the opposite direction as one would expect if the azimuthal force were generated by the Galactic Bar, and may indicate that forces caused by spiral arms or other overdensities are larger than those caused by the bar locally.

The cylindrical radial acceleration profile is consistent with a circular velocity curve of $\sim$\SI{237}{km/s}, in good agreement with the literature \citep{Eilers19, Zhou22, Poder23, Ou24}. We see no statistically significant gradient in the circular velocity curve within the volume of interest, arising probably from a mixture of the stationarity condition not holding and model uncertainties. The simple axisymmetric model is not able to capture a constant circular velocity curve because it is made up of components that do not have a constant circular velocity built in. We also draw comparisons to the line-of-sight accelerations from binary pulsar systems provided by \cite{Moran24} in appendix~\ref{appendix:pulsars}. While the accelerations from Deep Potential are consistent with measurements from binary pulsars, the pulsar measurements have much larger uncertainties, and do not presently serve as a strong test of our results. This can be expected to change in the coming years, as uncertainties on the pulsar measurements decrease and other direct acceleration measurements become available.

\begin{table}[]
\centering
\begin{tabular}{c|l}
Parameter & \multicolumn{1}{c}{This work} \\ \hline
% $\Omega_p\,(\si{km.s^{-1}.kpc^{-1}})$ & $\num{28.2}\pm \num{0.1}$    \\
% $a_{\odot,x}\,(\si{km/s/Myr})$ & $\num{7.02}\pm\num{0.03}$ \\
% $a_{\odot,y}\,(\si{km/s/Myr})$  & $\num{-0.01}\pm\num{0.03}$ \\
% $a_{\odot,z}\,(\si{km/s/Myr})$ & $\num{-0.042}\pm\num{0.007}$ \\
% $\rho(R=R_\odot)\,(\si{M_\odot.pc^{-3}})$ & $\num{0.086}\pm\num{0.010}$   \\
% $\rho_\mathrm{DM}\,(\si{M_\odot.pc^{-3}})$  & $\num{0.007}\pm\num{0.011}$\\
%
$\Omega_p$ & $\hphantom{-}\num{28.2} \hphantom{0}\pm\hphantom{0} \num{0.1}\hphantom{0} \ \si{km.s^{-1}.kpc^{-1}}$    \\
$a_{\odot,x}$ & $\hphantom{-}\num{7.02} \hphantom{0}\pm\hphantom{0} \num{0.03} \ \si{km/s/Myr}$ \\
$a_{\odot,y}$  & $\num{-0.01} \hphantom{0}\pm\hphantom{0} \num{0.03} \ \si{km/s/Myr}$ \\
$a_{\odot,z}$ & $\num{-0.042}\pm\num{0.007} \ \si{km/s/Myr}$ \\
$\rho(R=R_\odot)$ & $\hphantom{-}\num{0.086}\pm\num{0.010} \ \si{M_\odot.pc^{-3}}$   \\
$\rho_b(R=R_\odot)$ & $\hphantom{-}\num{0.079}\pm\num{0.007} \ \si{M_\odot.pc^{-3}}$   \\
$\rho_\mathrm{DM}$  & $\hphantom{-}\num{0.007}\pm\num{0.011} \ \si{M_\odot.pc^{-3}}$\\
\end{tabular}
\caption{Summary of the key results of our model, along with $1\sigma$ uncertainties. The volumes used for averaging the quantities are highlighted in Appendix~\ref{appendix:uncertainties}.}
\label{tab:fit_values}
\end{table}

\subsection{Pattern speed}
We compute a local pattern speed of $\Omega_p = \SI{28.2\pm 0.10}{km/s/kpc}$ within the volume of interest. The pattern speed should be interpreted as the rotation speed at which the local \SI{1}{kpc} of the Galaxy would appear stationary. It does not necessarily imply that the gravitational potential is rotating with the pattern speed, or that the gravitational potential is time-dependent at all in any reference frame. Rather, we are mapping the instantaneous value of the gravitational potential at the current moment in time. What we naturally expect is that different populations of tracers give different pattern speeds due to them experiencing the Galactic potential for different durations and in different regimes. We further expect there to be no one fixed pattern speed that is valid throughout the Galaxy. For example, regions dominated by the bar may have a different pattern speed from regions where the non-axisymmetric part of the potential is dominated by spiral arms. It is likely that the pattern speed we measure cannot be mapped to a single feature, such as the spiral arms or the Galactic Bar, but rather that it can be interpreted as some combination of said features in the solar neighborhood. We draw comparisons to literature estimates for the Galactic Bar and spiral arm pattern speeds in Figure~\ref{fig:omega_literature}. As expected, the value lies somewhere between literature estimates for the Galactic Bar and spiral arms. The Galactic Bar is faster-moving and extends out to $\sim \SI{5}{kpc}$ from the Galactic Center, but its influence, including resonances, is felt well into the solar neighborhood. The spiral arms are slower-moving and, due to difficulties in measuring their properties, there are multiple competing models for their properties. In general, the picture is that of flocculent spiral arms with some combination of two- and three-armed patterns and speeds ranging from \SI{12}{km/s} all the way up to \SI{33}{km/s}. Our value for the pattern speed implies that the corotation radius is very close to the solar location: $R_c = v_\mathrm{circ}/\Omega_p = \SI{8.4}{kpc} = \num{1.02}{R_0}$.

\subsection{Matter density}

We can compute the matter density $\rho = \nabla^2\rho / (4\pi G)$ by further differentiating the acceleration. We plot the vertical density profile in Figure \ref{fig:all_rho_z} and various two-dimensional projections in Figure \ref{fig:rho_2d}.
%and a two-dimensional slice in the $x-y$ plane integrated between $-\SI{0.25}{kpc} < z < \SI{0.25}{kpc}$.
The estimates of the matter density are generally noisier than those of the accelerations, as they are based on the derivatives of an already noisy quantity, especially because we do not apply any spatial smoothing on the resulting fields. In the Galactic midplane, we see fluctuations on the order of \SI{50}{\%}, which can be attributed to intrinsic spatial variation in the matter density, systematics from the modeling pipeline, and the stationarity condition not holding. We compute the average matter density at the solar Galactocentric radius using a vertically thin volume defined by $|R-R_0| < \SI{0.2}{kpc}, \SI{0.1}{kpc} < r < \SI{0.8}{kpc}, |z| < \SI{0.05}{kpc}$. Within this volume, we find $\rho(R=R_0) = \SI{0.086\pm0.010}{M_\odot/pc^3}$. We can compute the local dark matter density by subtracting the contribution of baryonic matter from the total matter density. We adapt the \citet{McKee15} model for the Baryonic density -- in the same way as is done in \cite{Lim2023} -- to compute the average baryonic matter density in the aforementioned volume to be $\rho_b(R\!=\!R_\odot) = \num{0.079}\pm\SI{0.007}{M_\odot.pc^{-3}}$. This implies a residual dark matter density of $\rho_\mathrm{DM}(R\!=\!R_\odot) = \SI{0.007\pm0.011}{M_\odot/pc^3}$. This is smaller than the uncertainties on our total density, and thus does not represent a statistically significant measurement of the local dark matter density.

\subsection{Uncertainties}

As described in Section~\ref{sec:uncertainty} and Appendix~\ref{appendix:uncertainties}, we estimate the contribution of different sources of error to the uncertainties on each quantity of interest (acceleration components, density, and pattern speed). We find that different sources of error contribute different proportions of the error budget for each quantity. Observational noise, shot noise (bootstrap errors), and normalizing flow random seed play an important role for all quantities of interest. It should be noted, however, that the attempts to quantify observational noise and shot noise are affected by variations arising from the random seed of the NFs, and that it is not obvious how much of the variation can be attributed to observational/shot noise and how much to NF variation. The errors stemming from the NF random seed could be reduced by increasing the number of flows we train, or through methodological improvements to the normalizing flow architecture. Uncertainties in $\vec x_0$ and $\vec v_0$ have little effect on $\rho(R=R_0)$ but have a much larger effect on inferred accelerations.
Interestingly, we find that increasing the shot noise and observational errors biases some measured parameters (in particular, $\Omega_p$ and $\rho$). For a more in-depth breakdown of the contributions of various sources of error, refer to Appendix~\ref{appendix:uncertainties}.

While not the dominant contributor to final uncertainties, it is possible to reduce the effect of observational errors by denoising the distribution function, utilizing our knowledge of the uncertainties on each individual star's measured position and velocity. \cite{Yan2025} explores this using normalizing flows and could serve as a drop-in addition after the training of the distribution function, though this would incur further computational costs.

It is also important to highlight the importance of the stationarity condition in the final errors. While the mock simulations in \citet{Kalda24} also exhibited $\sim\SI{50}{\%}$ errors in density, stemming from non-stationarity and model systematics, it is not straightforward to generalize the length scale and amplitude of those fluctuations to the \Gaia sample used in this work. The number of available particles in the mock simulation in \citet{Kalda24} is orders of magnitude smaller than is available from \Gaia in the same volume.

\begin{figure}
    \centering
    \includegraphics[width=\linewidth]{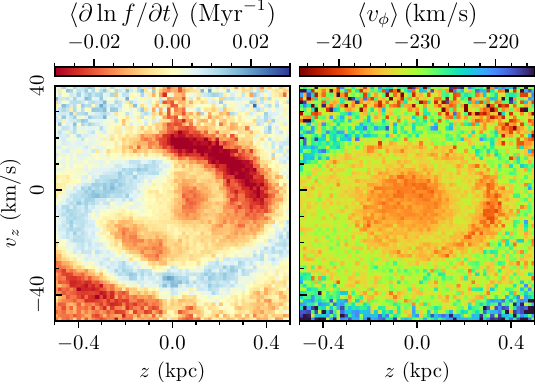}
    \caption{Imprint of the phase spiral in the $z-v_z$ projection in the residual non-stationarity $\left<\partial\ln f/\partial t\right>$ left over by the model (left panel) and a common tracer, $\left<v_\Phi\right>$, used to highlight the phase spiral (right panel). The residual non-stationarity can be interpreted as the inverse of the characteristic timescale over which the phase-space density at a specific volume undergoes significant changes. Both panels are averaged over a volume near the Sun (averaged within $|x|, |z|, |y| < \SI{500}{pc}$).}
    \label{fig:z_vz_dlnf_dt}
\end{figure}

\subsection{Non-stationarities}

Though Deep Potential minimizes non-stationarity in the system, there is, in general, no gravitational potential that renders a system fully stationary. After determining the potential, we can calculate the bulk motions predicted by the distribution function and potential. At each point in phase space, we calculate $\left(\pdv*{\ln f}{t}\right)_\Omega$, a measure of the inverse timescale over which the distribution function changes appreciably. Figure~\ref{fig:hist1d_dlnf_dt} shows the histogram of $\left(\pdv*{\ln f}{t}\right)_\Omega$ near the Sun, with a comparison drawn to a gravitational potential that has been trained with zero rotation speed (\textit{i.e.}, minimizing non-stationarity in the lab frame). As can be seen, the DF shows non-stationaries on characteristic timescales of \SI{\sim 5}{Myr} and \SI{\sim 20}{Myr} for the rotating and non-rotating models, respectively. The characteristic timescale for non-stationarity is similar to what was found by \cite{Lim2023}. Notably, however, they used a more luminous set of tracers and found the non-stationarity in the non-rotating case to be consistent with a timescale of \SI{5}{Myr}.

We have additionally investigated the possible origins of the non-stationarity in our model by correlating it with known dynamical features. To do this, we have examined its projection in the $v_x-v_y$ and $z-v_z$ planes near the Sun -- regions where structures such as moving groups and the phase spiral are prominent. In the $v_x-v_y$ plane, we did not find significant overlap with the moving groups. However, the $z-v_z$ plane, when averaged within the solar neighborhood, prominently highlights the phase spiral (see Figure~\ref{fig:z_vz_dlnf_dt}, serving as a sanity check that the model struggles with capturing non-stationary features, and that it can be used to identify real non-stationarities in the Milky Way.

\section{Conclusion}\label{sec:conclusion}

In this paper, we have applied the Deep Potential method to a \SI{1}{kpc} volume centered on the Sun, using a sample of \num{5.6} million upper-main-sequence stars with six-dimensional phase-space measurements from \Gaia DR3. The method makes use of a normalizing flow to model the full six-dimensional phase-space distribution of stars and a neural network to model the gravitational potential and further determines the local pattern speed that best renders the sample of stars stationary. As the approach demands a spatially uniformly complete sample, we have developed a smooth model of the observational completeness of our stellar sample, implemented using a neural network.

The method is able to resolve deviations from axisymmetry in the form of a systematic azimuthal acceleration in the order of \SI{0.05}{km/s/Myr} counter to the rotation of the Galaxy and to recover one-dimensional profiles in acceleration and density that are in broad agreement with the literature. We find a local pattern speed of $\Omega_p=\SI{28.2\pm0.1}{km/s/kpc}$, and total and dark matter densities at a solar distance from the Galactic Center of $\rho =\SI{0.086\pm0.010}{M_\odot/pc^3}$ and $\rho_\mathrm{DM}=\SI{0.007\pm0.011}{M_\odot/pc^3}$, respectively. Based on the radial acceleration profile, we find a local circular velocity of $v_\mathrm{circ}=\SI{237}{km/s}$, with only minor variation (as a function of radius) found within the volume of interest.

There are several straightforward directions to build on the work in this paper and to improve the underlying Deep Potential method. Use of multiple stellar populations with different luminosities would allow us to map the gravitational potential -- and thus the matter density -- throughout a larger volume of the Milky Way. With additional computational expense, the distribution function can also be de-noised by taking into account the reported \Gaia measurement errors. Additionally, one can modify the stationarity condition to capture differential rotation at different Galactocentric radii in the Milky Way disk. It is possible to leverage the overconstrained nature of the collisionless Boltzmann equation to fit corrections to our model of the selection function, similarly to \citet{Putney2024}. Newer normalizing flow architectures could help reduce the computational cost of modeling the distribution function. With these improvements, it should be possible to map the gravitational potential throughout a larger volume of the Galaxy and higher fidelity. Finally, it is possible to investigate the effect of non-stationarities and resonances on the recovered potential by applying Deep Potential to high-resolution zoom-in simulations of Milky-Way-like galaxies, such as Auriga \citep{Auriga24} and FIRE-2 \citep{Fire223}.

\begin{acknowledgments}
This work was supported by funding from the Alexander von Humboldt Foundation, through Gregory M. Green’s Sofja Kovalevskaja Award, and made use of the HPC systems Raven and Vera at the Max Planck Computing and Data Facility.

This work has made use of data from the European Space Agency (ESA) mission {\it Gaia} (\url{https://www.cosmos.esa.int/gaia}), processed by the {\it Gaia} Data Processing and Analysis Consortium (DPAC, \url{https://www.cosmos.esa.int/web/gaia/dpac/consortium}). Funding for the DPAC has been provided by national institutions, in particular, the institutions participating in the {\it Gaia} Multilateral Agreement. The authors thank Hans-Walter Rix, Soumavo Ghosh, David Hogg, Yuan-Sen Ting, and Lukas Eisert for helpful discussions on the method and manuscript.

Software citation information aggregated using \texttt{\href{https://www.tomwagg.com/software-citation-station/}{The Software Citation Station}} \citep{software-citation-station-paper, software-citation-station-zenodo}.

\software{\texttt{astropy} \citep{astropy:2013, astropy:2018, astropy:2022}, \texttt{Jupyter} \citep{2007CSE.....9c..21P, kluyver2016jupyter}, \texttt{matplotlib} \citep{Hunter:2007}, \texttt{numpy} \citep{numpy}, \texttt{python} \citep{python}, \texttt{scipy} \citep{2020SciPy-NMeth, scipy_15366870}, \texttt{scikit-learn} \citep{scikit-learn, sklearn_api, scikit-learn_14627164}, \texttt{Cython} \citep{cython:2011}, \texttt{dustmaps} \citep{2018JOSS....3..695M, dustmaps_10517733}, \texttt{gala} \citep{gala_JOSS, gala_13377376}, \texttt{galpy} \citep{2015ApJS..216...29B}, \texttt{h5py} \citep{collette_python_hdf5_2014, h5py_7560547}, \texttt{tensorflow} \citep{tensorflow_15009305}, \texttt{tqdm} \citep{tqdm_14231923}, \texttt{CMasher} \citep{2020JOSS....5.2004V,CMasher_14186007}, \texttt{healpy}, HEALPix package\footnote{http://healpix.sourceforge.net} \citep{Zonca2019, 2005ApJ...622..759G,healpy_15091710}}
\end{acknowledgments}

\appendix
\section{Feature importance for the model for the selection function}
\label{appendix:feature_importance}

We list the feature importance of the trained model for the selection function of our subset of stars in GDR3, as outlined in \ref{sec:sel}, in Table \ref{tab:feature_importance}. The feature importance is calculated using permutation feature importance \citep{Fisher2019VariableImportance}. The most important features are the color of a star, its apparent magnitude, and extinction. 
Perhaps surprisingly, the color of a star $G - G_\mathrm{RP}$ is more important than the apparent magnitude. The Gaia DR3 RVS selection function is sensitive to color because redder stars emit more flux in the RVS band, increasing their likelihood of passing the signal-to-noise and quality thresholds used for inclusion. The extinction plays a role because it not only changes the color of a star by reddening, but it can also increase detection efficiency by obscuring background stars and, therefore, reducing crowding. $\texttt{astrometric\_matched\_transits}$ corresponds to how many transits were used to compute astrometric solutions; the bigger the value, the higher the probability of passing internal cuts.
Finally, among the Cartesian components of the unit vector pointing towards the source, $z = \cos b\cos l$ contains the strongest signal due to correlating with dust clouds lying along the Galactic plane.

\begin{table}
\centering
\begin{tabular}{c|c}
$G-G_\mathrm{RP}$ & 1.1153 \\
$G$ & 0.7383 \\
$E$ & 0.7155 \\
$z$ & 0.2045 \\
$\texttt{astrometric\_matched\_transits}$ & 0.1805 \\
$y$ & 0.1780 \\
$x$ & 0.1279 \\
$\texttt{dec\_pmdec\_corr}$ & 0.1196 \\
$\texttt{ra\_dec\_corr}$ & 0.0786 \\
$\texttt{pmra\_pmdec\_corr}$ & 0.0769 \\
$\texttt{matched\_transits}$ & 0.0739 \\
$\texttt{visibility\_periods\_used}$ & 0.0685 \\
$\texttt{parallax\_pmdec\_corr}$ & 0.0578 \\
$\texttt{dec\_parallax\_corr}$ & 0.0398 \\
$\texttt{ra\_pmra\_corr}$ & 0.0314 \\
$\texttt{ra\_parallax\_corr}$ & 0.0313 \\
$\texttt{parallax\_pmra\_corr}$ & 0.0246 \\
$\texttt{astrometric\_params\_solved}$ & 0.0154 \\
$\texttt{ra\_pmdec\_corr}$ & 0.0140 \\
$\texttt{dec\_pmra\_corr}$ & 0.0123 \\
\end{tabular}
\caption{
Permutation importance of the features used for training the selection-function model. For each feature, we calculate the $R^2$ score, averaged over 10 permutations. The standard deviation of an individual feature importance is of the order \num{0.0001}. $E$ is derived from the \citet{Edenhofer23} dust map. $x = \cos b \cos l$, $y = \cos b\sin l$, and $z = \sin b$ are the Cartesian components of the unit vector pointing towards the source in Galactic coordinates. $\texttt{astrometric\_matched\_transits}, \texttt{matched\_transits}, \texttt{visibility\_periods\_used}, \texttt{astrometric\_params\_solved}$ are features pertaining to the scanning pattern of Gaia. Features with the suffix $\texttt{\_corr}$ describe the correlation between pairs of the five primary astrometric parameters measured by \Gaia.}
\label{tab:feature_importance}
\end{table}

\section{Comparisons with Pulsar acceleration}
\label{appendix:pulsars}

\begin{figure}
    \centering
    \includegraphics[width=0.5\linewidth]{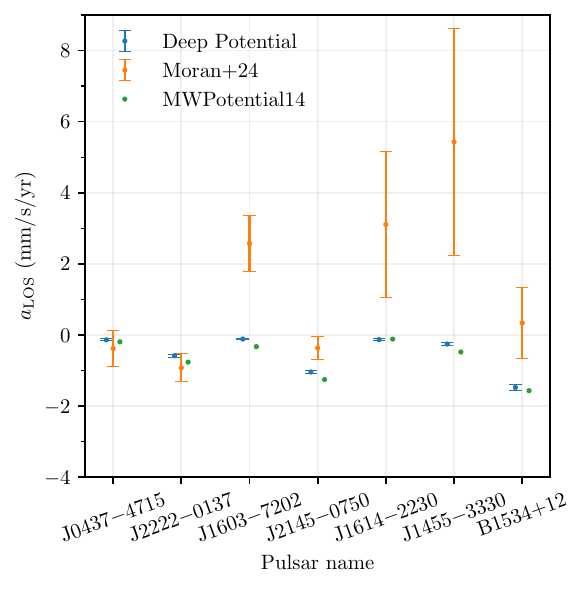}
    \caption{Comparisons between line-of-sight accelerations $a_{\mathrm{LOS}} \equiv (\vec a - \vec a_{\odot})\cdot \hat{r}$ measured using binary pulsars \citep{Moran24}, Deep Potential (this work) and \texttt{MWPotential2014} \citep{2015ApJS..216...29B}. For binary pulsars and Deep Potential, we show $1\sigma$ uncertainties.}
    \label{fig:pulsar_comparison}
\end{figure}

Out of all-sky direct acceleration measurements, only binary pulsar systems have provided a three-dimensional view of the line-of-sight accelerations in the Milky Way \citep{Donlon24, Moran24}. They rely on the principle of using pulsars as precise astrophysical clocks for measuring the orbital period of the binary and using general relativity to infer the acceleration from that. The resulting accelerations are relatively noisy but virtually model-free. As such, they can serve as unbiased validation for different models of the Galactic Potential. In Figure \ref{fig:pulsar_comparison} we draw comparisons between the line-of-sight accelerations from pulsars within \SI{1}{kpc} of the Sun \citep{Moran24}, Deep Potential, and \texttt{MWPotential2014} \citep{2015ApJS..216...29B}. In general, the uncertainties from the binary pulsars are a bit too high to draw significant conclusions about model performance, but all the pulsars fall reasonably within the reported uncertainties, and we can observe slightly better agreement from Deep Potential when compared to \texttt{MWPotential2014}.

\section{Quantifying uncertainties}
\label{appendix:uncertainties}
\begin{figure}
    \centering
    \includegraphics[width=\linewidth]{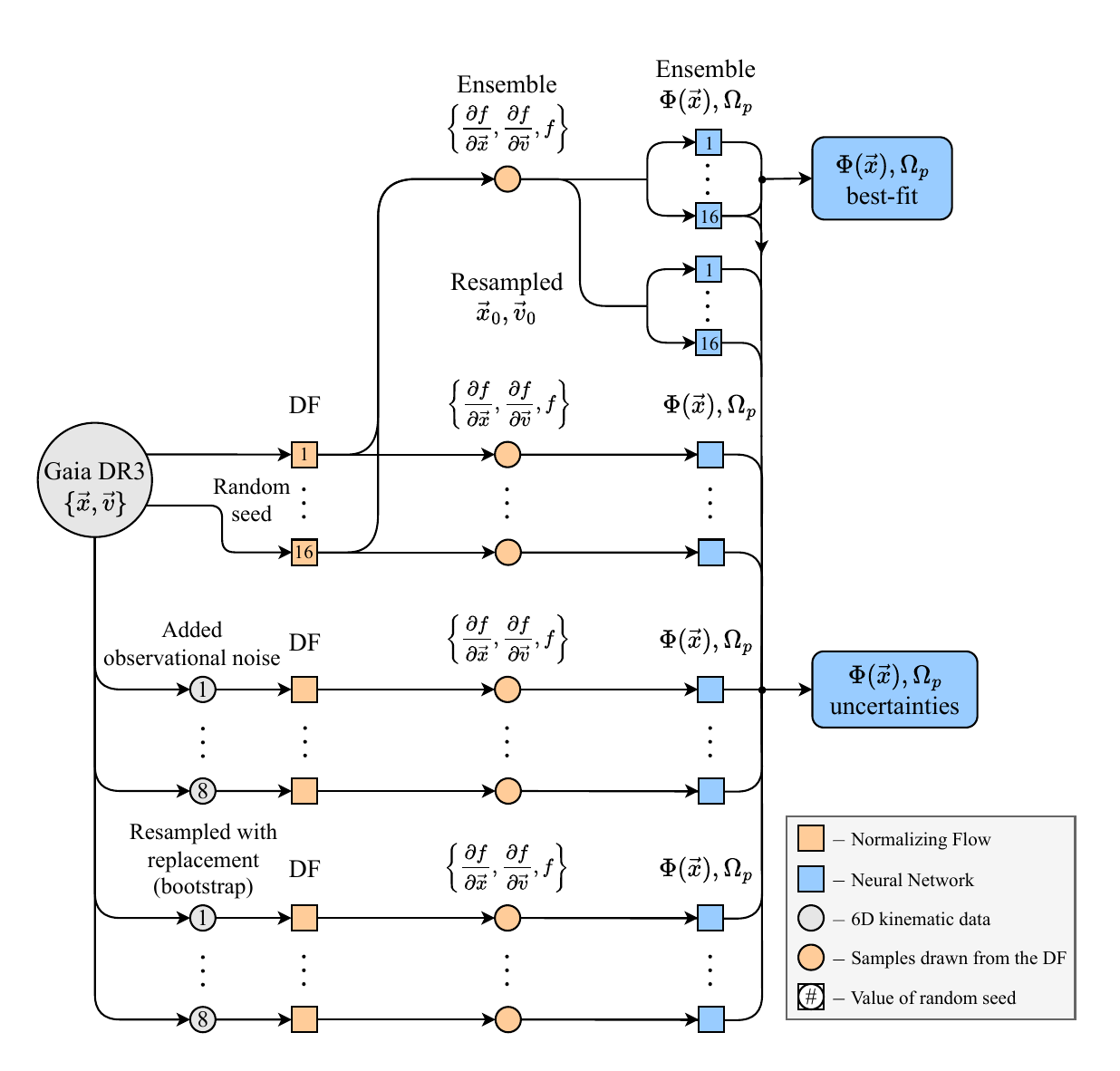}
    \caption{Illustration of the various models trained for Deep Potential, which are used to both determine a best estimate of the potential and to quantify its uncertainties. Squares indicate models, circles indicate datasets, nodes colored orange pertain to the distribution function, and nodes colored blue pertain to the gravitational potential and pattern speed $\Omega_p$.
    To obtain the best-fit value for the gravitational potential and $\Omega_p$, we first train 16 normalizing flows (with different random seeds) to approximate the distribution function. We obtain a best estimate of the distribution function and its gradients by taking the median of the 16 flows and of their gradients (Ensemble $\{\pdv*{f}{\vec x}, \pdv*{f}{\vec v}, f\})$. We use this best-fit distribution function to train 16 neural networks and pattern speeds (with different random seeds). The best-fit pattern speed and potential (or its derivatives) are defined by the median of these 16 pattern speeds and neural networks (or of their derivatives).
    We additionally estimate various sources of errors in the potential. For example, to quantify the effect of observational noise, we train 8 distribution functions on Gaia data with added noise. For each distribution function, we train a potential. From this ensemble, we can calculate variance in the inferred pattern speed, potential and derived quantities.
    %
    %On that sample, 16 different realizations of a neural network and $\Omega_p$ are trained, and the median is taken to get the final estimate.
    %
    From top to bottom on the figure, the following uncertainties are quantified: observational errors in $\vec x_0$ and $\vec v_0$ by resampling the respective quantities based on variations in the literature values, fluctuations in the random seed and convergence of the normalizing flow and neural network architecture, observational errors in the source data by resampling the 6D positions of stars based on their quoted errors, and finally, statistical uncertainties from the finite size of the dataset using bootstrap on the training data.}
    \label{fig:dpot_diagram}
\end{figure}

\begin{figure}
    \centering
    \includegraphics[width=\linewidth]{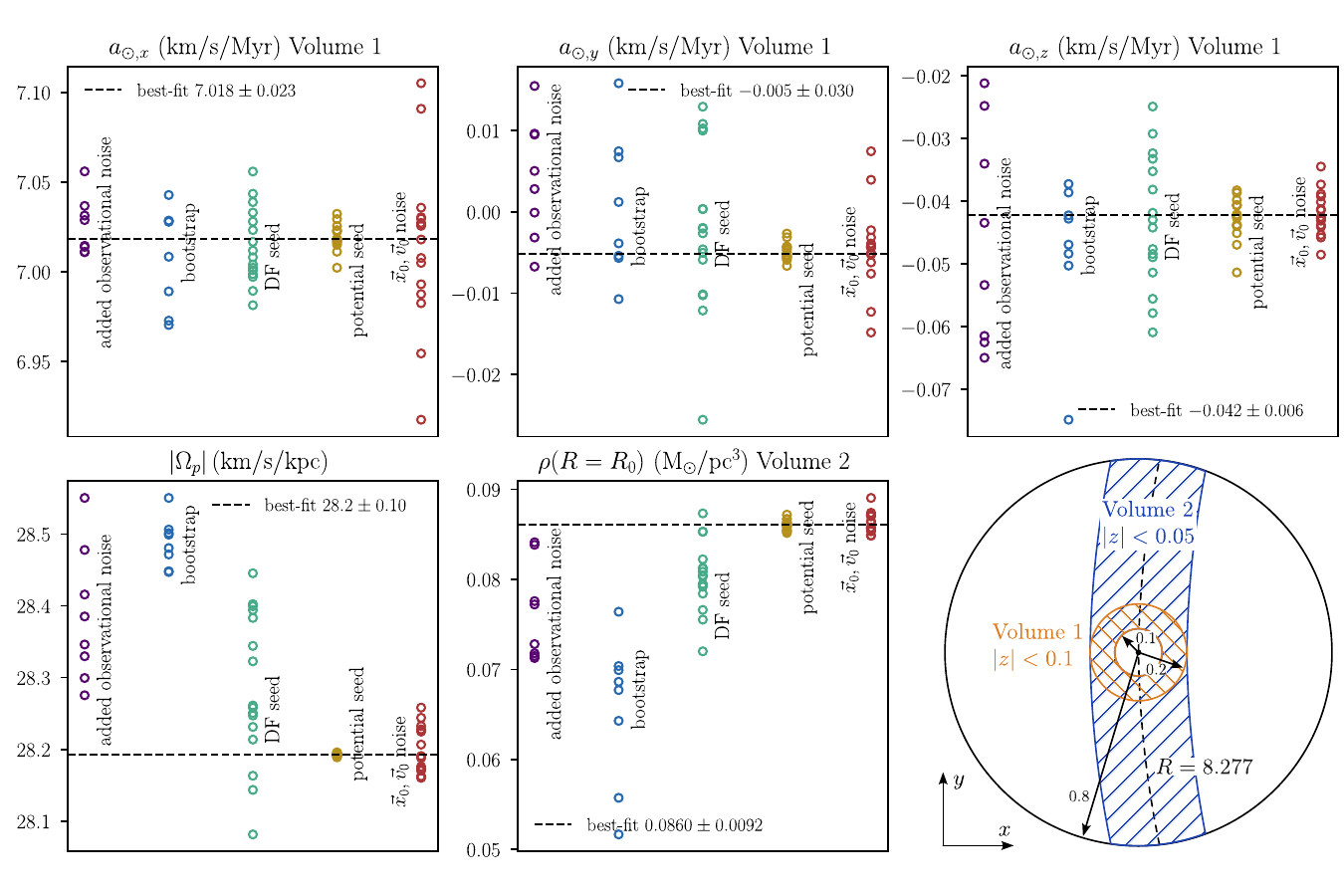}
    \caption{Distribution of scalar value quantities from different uncertainty estimation pipelines visualized in Figure \ref{fig:dpot_diagram}. The scalar quantities are: the average acceleration in the solar neighborhood $a_{\odot,x}, a_{\odot,y}, a_{\odot,z}$, the pattern speed of the solar neighborhood within \SI{1}{kpc} from the Sun, and the average total matter density $\rho(R=R_0)$ at the solar distance $R = R_0$ from the Galactic Center. The bottom right diagram visualizes which volumes the quantities are averaged over. The central \SI{0.1}{kpc} and outer \SI{0.8}{kpc} are excluded due to a lack of data and systematics from edge effects, respectively. In each subplot, the columns represent the spread of values due to the following five effects: adding observational noise to the observed 6D kinematics of the training data (``added observational noise''), resampling the stars in the training data using bootstrap (``bootstrap''), variations in the starting seed of the normalizing flows representing the DF (``DF seed''), variations in the starting seed of the neural network representing the gravitational potential while using the best-fit DF (``potential seed''), and adding observational noise to the literature values of $\vec x_0, \vec v_0$ while using the best-fit DF (``$\vec x_0, \vec v_0$ noise''). The best-fit value is calculated using the median of ``potential seed''. For discussion, see \ref{appendix:uncertainties}}
    \label{fig:scalar_values}
\end{figure}

A visualization of the full modeling pipeline, including error estimates, is provided in Figure \ref{fig:dpot_diagram}. Because Deep Potential is not probabilistic nor does it employ forward modeling, it does not provide posterior distributions or formal uncertainty quantification. To estimate uncertainties, a frequentist, ensemble-based approach is employed instead. For some of the uncertainty analysis, the models trained for the best-fit estimate can be repurposed. To reiterate, the best-fit estimate is obtained by first training 16 different normalizing flows to represent the distribution function, and then by taking the median of their density and gradients. Then, 16 different neural networks are trained for the gravitational potential, and their median values are calculated for the final $\Omega_p$ and $\Phi(\vec x)$. The various uncertainties and their estimation procedures are given as follows:

\begin{itemize}
    \item To emulate observational errors, we generate eight different realizations of the training data by resampling their 6D astrometry ($\alpha, \delta, \varpi, \mu_\alpha, \mu_\beta, v_r$) based on their quoted errors and assuming that they are normally distributed. For each realization, one NF and one NN are trained.
    \item To estimate the Poisson noise arising from the finite size of the dataset, we use bootstrap to resample the training data with replacement a total of eight different times and, once again, train a single NF and NN for each realization.
    \item To estimate the variations in the convergence of the normalizing flows used to represent the distribution function due to their varying random seeds, we use the 16 NFs used for the best-fit estimate and train a single NN on each one of them separately.
    \item To estimate the variations in the convergence of the neural networks used to represent the gravitational potential due to their varying random seeds, we use the 16 NNs trained on the best-fit DF sample and observe the variation from model to model.
    \item To estimate the effects from the assumed literature values of the distance to the Galactic Center and its velocity with respect to the Sun, we train 16 different NNs on the best-fit DF by resampling the literature values of $\vec x_0, \vec v_0$ using their quoted errors and by assuming them to be normally distributed. We assume the errors do not have any covariance when expressed in terms of the proper motion of Sgr A*, its parallax, and radial velocity.
\end{itemize}

One of the challenges is that it is difficult to truly disentangle the various effects from each other. Notably, variations arising from the seed of the normalizing flow and neural networks affect every previously mentioned source of error. It is possible, though, to reduce their effect. For example, as is done for the best-fit model, by averaging multiple flows and neural networks, one reduces the effect of the variations coming from model convergence. Assuming the variations are normally distributed, we would expect the averaging to reduce the variance 16 times. While our empirical tests favor this interpretation, mathematically, the averaging of models is not guaranteed to behave that way, nor result in the convergence happening towards the true underlying gravitational potential. This leads to another challenge, which is that some of these sources of error also result in systematic shifts in the inferred $\Phi(\vec x)$ and $\Omega_p$.

This is most readily observed by looking at specific scalar values and how they vary. Figure \ref{fig:scalar_values} shows how the average acceleration in the solar neighborhood $a_{\odot,x}, a_{\odot,y}, a_{\odot,z}$, $\Omega_p$, and the density at solar distance from the Galactic Center $\rho(R=R_0)$ vary due to the previously mentioned effects. Firstly, we see that the dominating sources of error are from observational errors, shot noise (bootstrap errors), DF convergence and the literature values of $\vec x_0$ and $\vec v_0$. Secondly, it is apparent that observational errors and bootstrapping serve to give systematic shifts in the inferred quantities. This means that the best-fit values for $\Phi(\vec x), \Omega_p$ have leftover systematics from observational errors. Quantifying the magnitude of this warrants further study. Thirdly, we can see that the various effects do not commute. For example, we see that the spread in the values for bootstrap and added observational noise gets reduced when compared to the variations arising from model random seeds only, even though the random seed of the model ought to affect them all equally. This is not too surprising as, for instance, adding observational noise to the training data smooths the distribution function over some length scale, making it easier for the normalizing flow to approximate and, hence, converge toward the true underlying distribution.

Taking all of this together, we estimate the errors by combining the effects of adding observational noise, bootstrapping, model random seed variation, and $\vec x_0, \vec v_0$ resampling. We add the effects in quadrature to get a conservative estimate for the error. The true error from these effects is smaller, as in reality they are not independent of each other, but we do this to avoid underestimating the error. We also point out that this is not an exhaustive exploration of sources of errors, for example, systematic offsets coming from the choice of normalizing flow and neural network architecture and the stationarity assumption not holding true.


%% the collaboration and author truncation commands are used. It has to

%%\listofchanges

\bibliography{main}{}
\bibliographystyle{aasjournal}

\end{document}